\documentclass[11pt,french,english,aps,prl,preprint,nofootinbib]{revtex4}
\usepackage[T1]{fontenc}
\usepackage[latin9]{inputenc}
\usepackage[a4paper]{geometry}
\usepackage{color}
\usepackage{textcomp}
\usepackage{bm}
\usepackage{amsmath}
\usepackage{graphicx}

\makeatletter

\providecommand{\tabularnewline}{\\}

\@ifundefined{textcolor}{}
{%
 \definecolor{BLACK}{gray}{0}
 \definecolor{WHITE}{gray}{1}
 \definecolor{RED}{rgb}{1,0,0}
 \definecolor{GREEN}{rgb}{0,1,0}
 \definecolor{BLUE}{rgb}{0,0,1}
 \definecolor{CYAN}{cmyk}{1,0,0,0}
 \definecolor{MAGENTA}{cmyk}{0,1,0,0}
 \definecolor{YELLOW}{cmyk}{0,0,1,0}
}

\usepackage{bbm}
\usepackage{braket}

\makeatother

\usepackage{babel}
\makeatletter
\addto\extrasfrench{%
   \providecommand{\fg}{\ifdim\lastskip>\z@\unskip\fi~\frqq}%
}

\makeatother
\begin{document}

\title{A Molecular Density Functional Theory Approach to Electron Transfer
Reactions.}

\author{Guillaume Jeanmairet}

\affiliation{Sorbonne Universit\'e, CNRS, Physico-Chimie des \'Electrolytes et Nanosyst\`emes
Interfaciaux, PHENIX, F-75005 Paris, France}

\affiliation{R\'eseau sur le Stockage \'Electrochimique de l'\'Energie
(RS2E), FR CNRS 3459, 80039 Amiens Cedex, France}

\author{Benjamin Rotenberg}

\affiliation{Sorbonne Universit\'e, CNRS, Physico-Chimie des \'Electrolytes et Nanosyst\`emes
Interfaciaux, PHENIX, F-75005 Paris, France}

\affiliation{R\'eseau sur le Stockage \'Electrochimique de l'\'Energie
(RS2E), FR CNRS 3459, 80039 Amiens Cedex, France}

\author{Maximilien Levesque}

\affiliation{PASTEUR, D\'epartement de chimie, \'Ecole normale sup\'erieure, PSL University,
Sorbonne Universit\'e, CNRS, 75005 Paris, France}

\author{Daniel Borgis}

\affiliation{PASTEUR, D\'epartement de chimie, \'Ecole normale sup\'erieure, PSL University,
Sorbonne Universit\'e, CNRS, 75005 Paris, France}

\affiliation{Maison de la Simulation, CEA, CNRS, Universit\'{e} Paris-Sud, UVSQ,
Universit\'{e} Paris-Saclay, 91191 Gif-sur-Yvette, France}

\author{Mathieu Salanne}

\affiliation{Sorbonne Universit\'e, CNRS, Physico-Chimie des \'Electrolytes et Nanosyst\`emes
Interfaciaux, PHENIX, F-75005 Paris, France}

\affiliation{R\'eseau sur le Stockage \'Electrochimique de l'\'Energie
(RS2E), FR CNRS 3459, 80039 Amiens Cedex, France}

\affiliation{Maison de la Simulation, CEA, CNRS, Universit\'{e} Paris-Sud, UVSQ,
Universit\'{e} Paris-Saclay, 91191 Gif-sur-Yvette, France}

\begin{abstract}
Beyond the dielectric continuum description initiated by Marcus theory,
the standard theoretical approach to study electron transfer
(ET) reactions in solution or at interfaces is to use classical force
field or\textit{ ab initio} Molecular Dynamics simulations. We present
here an alternative method based on liquid-state theory, namely molecular
density functional theory, which is numerically much more efficient
than simulations while still retaining the molecular nature of the
solvent. We begin by reformulating molecular ET theory in a density
functional language and show how to compute the various observables
characterizing ET reactions from an ensemble of density functional
minimizations. In particular, we define within that formulation the relevant
order parameter of the reaction, the so-called vertical energy gap,
and determine the Marcus free energy curves of both reactant and product
states along that coordinate. Important thermodynamic quantities such
as the reaction free energy and the reorganization free energies follow.
We assess the validity of the method by studying the model Cl$^{0}$
 $\rightarrow$ Cl$^{+}$ and Cl$^{0}$  $\rightarrow$  Cl$^{-}$
ET reactions in bulk water for which molecular dynamics results are
available. The anionic case is found to violate the standard Marcus
theory. Finally, we take advantage of the computational efficiency
of the method to study the influence of a solid-solvent interface on the ET, by
investigating the evolution of the reorganization free energy of the
Cl$^{0}$   $\rightarrow$ Cl$^{+}$ reaction when the atom approaches
an atomistically resolved wall.
\end{abstract}
\maketitle

\section{Introduction\label{sec:Intro}}

Electron transfer (ET) reactions play a central role in a wide range
of chemical systems, including energy storage and harvesting in electrochemical
devices or biological processes such as aerobic respiration and photosynthesis.
This ubiquity can explain the considerable amount of experimental,
theoretical and simulation studies that have been devoted to this
class of reactions \citep{marcus_electron_1997}. The widely accepted
theory of ET reaction in solution was proposed by Marcus \citep{marcus_electrostatic_1956,marcus_theory_1956,marcus_reorganization_1990}.
It is based on the description of the solvent by a dielectric continuum.
The macroscopic fluctuations of the solvent are represented by an out-of-equilibrium
polarization field, and the free energy is a functional depending
quadratically on this polarization. It eventually provides a simple
two-state picture, where the free energy of each state depends quadratically
on a reaction coordinate. This famous two-parabola picture has been
used with great success to interpret experimental results and to make
predictions \citep{miller_intramolecular_1984}. However, Marcus theory
does not take into account the molecular nature of the solvent which
can break the linear assumption of solvent response. In such cases,
we must resort to molecular simulation.

The vast majority of simulation studies on ET reaction have been carried out 
using Molecular Dynamics (MD). For example, the pioneering work of
Warshel demonstrated that the vertical energy gap is the appropriate
reaction coordinate \citep{hwang_microscopic_1987} and that the fluctuations
of this quantity are Gaussian. Such Gaussian statistics give rise
to the famous parabola picture of Marcus for the free energy profile.
A strict Gaussian behaviour is equivalent to a linear response of the
solvent to the field generated by the solute; it also implies that
the two free energy parabolas have the same curvature because the
solvent fluctuations are identical for the two states \citep{tachiya_relation_1989,tachiya_generalization_1993}.

This validity of the Gaussian assumption has been verified in numerous
studies since, for ET in solution \citep{kuharski_molecular_1988,blumberger_quantum_2006}
or in complex biological systems \citep{simonson_gaussian_2002,sterpone_linear_2003},
using either classical or \textit{ab initio }MD. However, there is
evidence that some systems do not obey the Marcus assumptions \textit{i.e.
}the free energy curves of the two states cannot be represented by
a pair of identical parabolas. There are several possible origins
of such a discrepancy \citep{lande_chapter_2016}, in particular the
fact that reactant and product may have quite different solvation
states. This can happen when the ET occurs between neutral and charged
states, as predicted by Kakitani and Mataga \citep{kakitani_new_1985,kakitani_different_1986,kakitani_comprehensive_1987}
and observed since in classical \citep{carter_solute-dependent_1989,li_confinement_2017,hartnig_molecular_2001}
and \textit{ab initio} \citep{blumberger_free_2008,vuilleumier_extension_2012}
simulations. Several extensions of Marcus theory have been put forward 
to take into account the various origins of non-linearity \citep{matyushov_modeling_2000,small_theory_2003,vuilleumier_extension_2012,jeanmairet_chapter_2013}. 

The investigation of ET reactions by MD is quite challenging since
it usually requires the computation of solvation free energies the free energies curves as a function of an order parameter which remains
a demanding task. Indeed, it necessitates a proper sampling of the solvent
configurations around the barrier. If the activation energy is high
it requires the use of biases such as umbrella sampling \citep{torrie_nonphysical_1977}
coupled with histogram analysis techniques to reconstruct the unbiased
data \citep{ferrenberg_optimized_1989,shirts_statistically_2008,tan_theory_2012}.
This typically implies to run simulations on half a dozen fictitious intermediate
 states to study a single system. 

To compute free energies, there exist alternative techniques based
on statistical theory of liquids, which offer the advantage of keeping
a molecular description of the solvent while avoiding to sample the
instantaneous microscopic degrees of freedom. Among the different
approaches one can mention integral equation theory either in its molecular
\citep{fries_solution_1985} or multiple sites formulation (RISM) \citep{chandler_optimized_1972,hirata_extended_1981}
and its 3D-RISM version \citep{kovalenko_three-dimensional_1998,imai_hydration_2006}.
Another method is the classical density functional theory (cDFT) of
liquids \citep{mermin_thermal_1965,evans_nature_1979} which describes
the response of a fluid in the presence of a perturbation by introducing
a functional of the fluid density. 
%This functional equals the grand
%potential at its minimum which is attained for the equilibrium fluid
%density.
Minimization of the functional yields the grand potential at equilibrium fluid density.
 Some of us have previously introduced the molecular density
functional theory (MDFT) \citep{ramirez_density_2002,ramirez_density_2005}  which is able to provide the solvation free energy
and the solvation structure of any solute embedded in a molecular
solvent described by its inhomogeneous density field. The solvent
density is a function of space coordinates and of the  orientation;
 hence the functional must be minimized on a 6D grid: 3 dimensions
for the cartesian coordinates and 3 dimensions for the three Euler
angles. This formalism can be used to solvate any simple or complex
solutes \citep{levesque_solvation_2012}. We proposed functionals
for several solvents \citep{borgis_molecular_2012,ramirez_direct_2005}
with particular attention paid to the case of water \citep{jeanmairet_molecular_2013-1,jeanmairet_molecular_2016}.
The most advanced version of the functional is equivalent to the molecular
Ornstein-Zernike theory supplemented by the hypernetted-chain closure
(HNC) \citep{ding_efficient_2017} for the solute-solvent correlations
and can be minimized efficiently thanks to the use of rotational invariants
in an optimal frame. The accuracy on the predictions of solvation
free energies is promising as illustrated on the FreeSolv database
\citep{luukkonen_high-throughput_2018}. We shall take advantage of this
accuracy to put forward an efficient way to compute the free energy curves.

An application of the MDFT formalism to ET reactions in acetonitrile
was proposed some years ago \citep{borgis_molecular_2012}. In this
article we extend this approach and apply it to ET in
aqueous solutions. In section \ref{sec:Theory}, after recalling some
basics of ET theory and giving a very short description of the MDFT
framework, we show how to compute the key quantities of ET reactions
with MDFT. In particular, we show that the average vertical energy
gap is an appropriate order parameter for the ET reaction. We prove
that for a given set of external potentials the free energy functional
is actually a function of this order parameter. We derive expressions
to compute the free energy curves (FEC) and the reorganization free
energies.

In section \ref{sec:Implementation} we first validate the framework
on the simplest solute in water, \textit{i.e. }a single neutral
or charged chlorine atom modeled by a Lennard-Jones site,  before
studying the influence of the presence of a solid-solvent interface on the reorganization free energy;
to this purpose we investigate the ET of this solute as a function of
its distance to an atomistically resolved wall.

%\subsection{This is the subsection heading style}
%Section headings can be typeset with and without numbers.\cite{Abernethy2003}

%\subsubsection{This is the subsubsection style.~~} These headings should end in a full point.  

%\paragraph{This is the next level heading.~~} For this level please use \texttt{\textbackslash paragraph}. These headings should also end in a full point.

\section{Theory\label{sec:Theory}}

\subsection{Electron Transfer Reaction}
\begin{figure}[]
\centering
  \includegraphics[width=0.4 \textwidth]{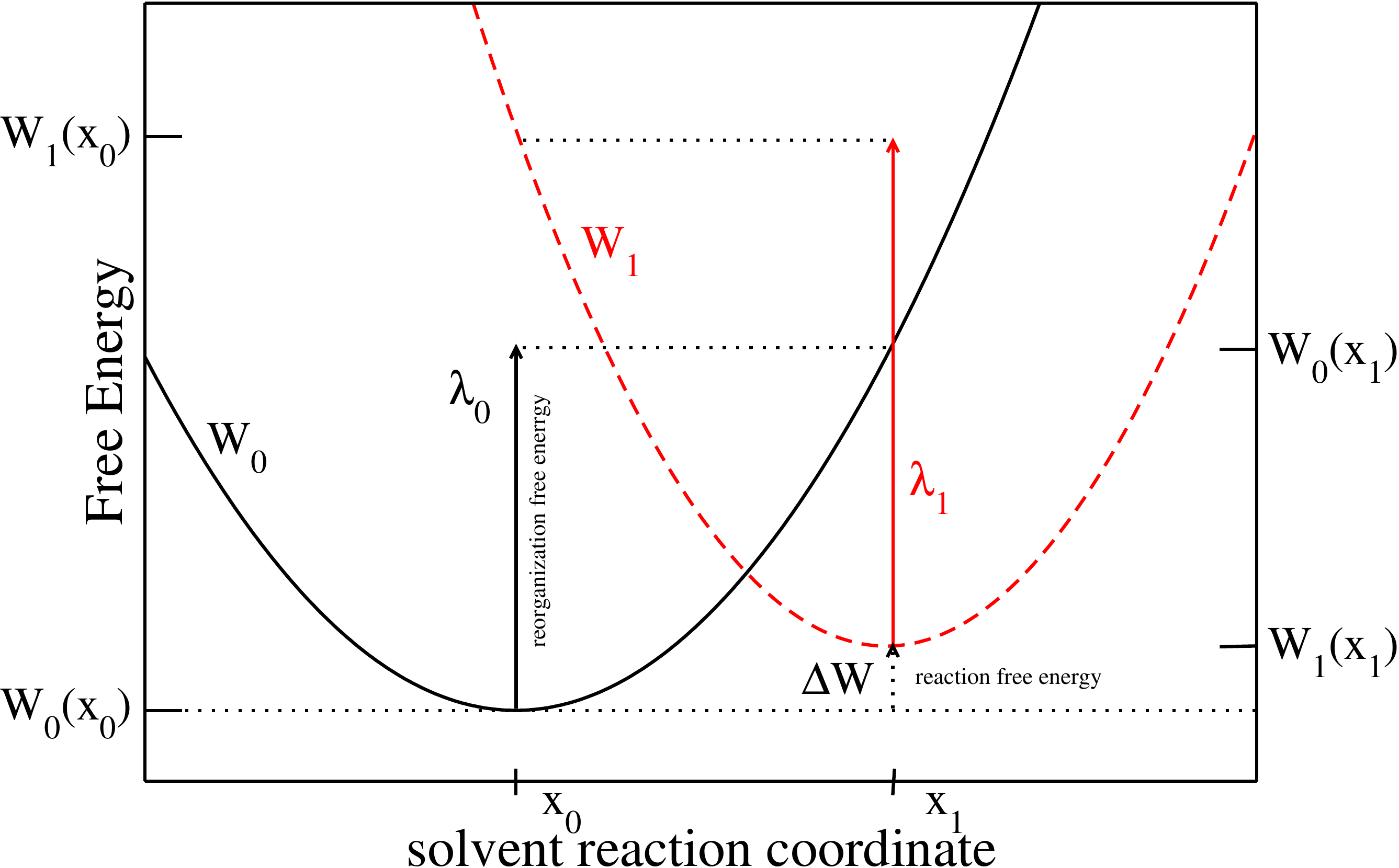}
  \caption{Schematic representation of a solvent controlled ET reaction. The
diabatic free energy curves for state 0 and 1 are represented in plain
black and dashed red, respectively. The reorganization free energies
$\lambda_{0}$ and $\lambda_{1}$ for the two states are represented
with full arrows, the reaction free energy with a dotted arrow.}
  \label{fig:Marcus_Plot}
\end{figure}

We limit ourselves to the study of ET reactions of solutes which interact
with the solvent through a classical force field. Moreover, the solutes
we consider in this article are rigid entities composed of a set of
Lennard-Jones sites and point charges. An ET reaction
involving two solutes of this type would correspond to an outer-sphere ET because there are no structural changes of the solutes. 
 This implies that the ET reaction
is completely controlled by the solvent response, as considered in
Marcus' original paper \citep{marcus_theory_1956}. The physics of
the system can be described by the two crossing free energy curves
of the system before (0) and after (1) the ET. A schematic view of
the two FEC is presented in Fig.\ \ref{fig:Marcus_Plot} where some
of the quantities necessary to describe the process are shown.
The order parameter $x$ describes the solvent configuration around
the solute, thus the abscissa $x_{0}$ of the minimum
of the FEC $W_{0}$ corresponds to a solvent in equilibrium with state
$0$. We emphasize that several microscopic solvent configurations
correspond to an identical value of the order parameter.

Values of the order parameter differing from $x_{0}$ correspond to
solvent configurations that are not in equilibrium with state $0$.
The more the solvent configuration differs from equilibrium,
the more the free energy increases. The difference between the minima
of the 2 FEC corresponds to the free energy difference between the
two states, each surrounded by a solvent in equilibrium with these states,  \textit{i.e} the reaction free energy, $\Delta W$. Two others
key quantities appear in Fig.\ \ref{fig:Marcus_Plot}: the reorganization
free energies $\lambda_{0}$ (resp. $\lambda_{1}$) which represent
the cost in free energy to solvate state $0$ (resp. $1$) in a solvent 
 in equilibrium with the other state. The difference
in free energy between the transition state (the crossing point) and
state $0$ controls the kinetics of the $0\rightarrow1$ reaction.

We emphasize that in the Marcus picture, the solvent is treated as
a continuum which responds linearly to the electric field generated
by the solute. This implies that the FEC of the 2 states are identical
parabolas. As a consequence, there is a unique reorganization free
energy parameter $\lambda=\lambda_{0}=\lambda_{1}$. The objective
of the present paper is to show how to compute the various quantities appearing
in Fig.\ \ref{fig:Marcus_Plot}, within MDFT. This could be a way to test
the validity of Marcus assumption when the molecular nature of the
solvent is taken into account, while taking advantage of the numerical
efficiency of MDFT compared to MD.

\subsection{Molecular Density Functional Theory}

We briefly recall  the fundamentals of MDFT which belong to the
more general class of cDFT. Based on the Hohenberg-Kohn ansatz \citep{hohenberg_inhomogeneous_1964},
Mermin introduced the framework of density functional theory (DFT)
at finite temperature for the inhomogeneous 
electron gas \citep{mermin_thermal_1965}. Later, Evans rewrote the
theory for a classical system, setting the foundations of cDFT
which describes the response of a fluid to an external perturbation
\citep{evans_nature_1979}. 
%cDFT states that in the grand-canonical
%ensemble it is possible to write a functional $\Theta$ of the fluid
%density $\rho$. At its minimum this functional is equal to the grand
%potential which is reached for the equilibrium fluid density.

MDFT is designed to study solvation problems. The fluid perturbed
by the presence of one or several solutes is described by its density
field. In the following we will always consider liquids, referred to 
as the solvent. It is advantageous to define a new functional $F$,
as the difference between the functional of the solvent in the presence
of the solute $\Theta$ and that of the homogeneous solvent,

\begin{equation}
F[\rho]=\Theta[\rho]-\Theta_{\text{H}}.\label{F=00003DF-F0}
\end{equation}

With this definition, the functional  at its minimum is equal to the solvation free
energy. Because both the solute and the solvent  are in most cases molecules with several atomic sites, their
interactions depend on both the  position of the centres of mass (CM) and orientations.
 Hence, the solvent density will be denoted by $\rho(\bm{r},\bm{\Omega})$
where $\bm{r}$ is the position in cartesian coordinates and $\bm{\Omega}$
the orientation described by three Euler angles ($\theta,\phi,\psi$).
The density minimizing the functional is the full equilibrium solvent
density around the solute which may be integrated to
recover the usual radial distribution functions.

The usual strategy  to have a workable expression of this functional
is to split it into the sum of ideal, excess and external contributions,
\begin{equation}
F[\rho]=F_{\text{id}}[\rho]+F_{\text{ext}}[\rho]+F_{\text{exc}}^{\text{exact}}[\rho].\label{eq:F=00003DFid+Fexc+Fext}
\end{equation}
In eq.\ref{eq:F=00003DFid+Fexc+Fext} the ideal term corresponds to
the entropy of the non-interacting fluid, which reads
\begin{equation}
F_{\text{id}}[\rho]=k_{B}T\iint\left[\rho\left(\bm{r},\bm{\Omega}\right)\ln\left(\frac{\rho\left(\bm{r},\bm{\Omega}\right)}{\rho_{\text{H}}}\right)-\rho\left(\bm{r},\bm{\Omega}\right)+\rho_{\text{H}}\right]d\bm{r}d\bm{\Omega},\label{eq:Fid}
\end{equation}
 where $k_{B}$ is the Boltzmann constant, $T$ the temperature in
Kelvin and $\rho_{\text{H}}=\frac{n_{\text{H}}}{8\pi^{2}}$ with $n_{\text{H}}\ $the
density of the homogeneous solvent. The second term in eq.\ref{eq:F=00003DFid+Fexc+Fext}
account for the perturbation by the solute. The solute acts on the solvent
via an "external" potential $V_{\text{ext}}$, typically the sum
of a Lennard-Jones term and of an electrostatic term. In the usual case of pair-wise additive interaction, $V_{\text{ext}}$ reads
\begin{equation}
V_{\text{ext}}(\bm{r},\bm{\Omega})=\sum_{i=\text{solvent site}}\sum_{j=\text{solute site}}v_{ij}(|\bm{r}+\bm{r_{i\bm{\Omega}}}-\bm{r}_j|)
\end{equation}
where $v_{ij}$ is the pair potential between site $i$ of solvent and site $j$ of the solute and $\bm{r_{i\bm{\Omega}}}$ denotes the position of site $i$ when the solvent molecule has the orientation $\bm{\Omega}$.
 The expression of the external functional is:
\begin{equation}
F_{\text{ext}}[\rho]=\iint\rho\left(\bm{r},\bm{\bm{\Omega}}\right)V_{\text{ext}}(\bm{r},\bm{\Omega})d\bm{r}d\bm{\Omega}.\label{eq:Fext}
\end{equation}
Finally, the last term corresponds to the solvent-solvent interactions.
%There are no practical expression for this functional and it is usually rewritten as
An exact expression for this term is available \cite{hansen_theory_2006} but in practice it is rewritten as the sum of two terms
\begin{equation}
F_{\text{exc}}^{\text{exact}}[\rho]=F_{\text{exc}}[\rho]+F_{b}[\rho],\label{eq:Fexc def}
\end{equation}
 where $F_{\text{exc}}$ is an approximation of the excess
functional. This defines the correction or "bridge" functional $F_{b}$ as the difference
between the exact functional and this approximation. To date, the
most advanced expression of $F_{\text{exc}}$ for water is  
that  recently  used by Ding and coworkers which corresponds
to the hypernetted chain approximation \citep{ding_efficient_2017}.
\begin{align}
&\beta F_{\text{exc}}[\rho]=\label{eq:FexcHNC}\\
&-\frac{1}{2}\iiiint\Delta\rho(\bm{r}_{1},\bm{\Omega}_{1})c(\left\Vert \bm{r}_{1}-\bm{r}_{2}\right\Vert ,\bm{\Omega}_{1},\bm{\Omega}_{2})\Delta\rho(\bm{r}_{2},\bm{\Omega}_{2})d\bm{r}_{1}d\bm{\Omega}_{1}d\bm{r}_{2}d\bm{\Omega}_{2} \nonumber 
\end{align}
 where $\beta=\left(k_{B}T\right)^{-1}$ and $c(\left\Vert \bm{r}_{1}-\bm{r}_{2}\right\Vert ,\bm{\Omega}_{1},\bm{\Omega}_{2})$
is the bulk direct correlation function while $\Delta\rho=\rho-\rho_{\text{H}}$.
$c$ depends on the distance between 2 solvent molecules and on their
relative orientation defined by six Euler angles instead of 4 in the case of 
acetonitrile \citep{borgis_molecular_2012}. The calculation
of the functional of eq.\ref{eq:FexcHNC} in this general case is
still feasible thanks to the efficient FFT algorithm \citep{frigo_design_2005}
to handle the spatial convolution and to the use of rotational invariants
to handle the angular one \citep{blum_invariant_1972-1,blum_invariant_1972,blum_invariant_1973}.

We have previously developed several approximation for the bridge
term $F_{b}$ \citep{levesque_scalar_2012,jeanmairet_molecular_2013,jeanmairet_molecular_2015}
but we shall neglect it in this paper and  refer to this functional
formulation as the HNC functional.
While the numerical problem involved in MD is the sampling of phase space, MDFT involves an optimization which is numerically more efficient.
Consequently,  to compute solvation free energies  of a spherical solute in water,  MDFT requires  $\approx 10$ cpu minutes with our lab-developed program while it requires  $\approx 100$ cpu hours to compute the same quantity with commercially available MD codes.
\subsection{ET reaction in the MDFT framework}

\subsubsection{Theory}

From a MDFT perspective, the two states 0 and 1 of the ET reaction
correspond to two functionals $F_{0}$ and $F_{1}$ differing only
by their external potentials $V_{0}$ and $V_{1}$ in eq.\ref{eq:Fext}.
If we denote by  $\rho_{0}$ and $\rho_{1}$ the equilibrium solvent densities
of states $0$ and $1$, obtained by minimization of $F_{0}$ and
$F_{1}$,  the reaction free energy can be expressed as
\begin{equation}
\Delta W=F_{1}[\rho_{1}]-F_{0}[\rho_{0}]+\Delta E_{0}.\label{eq:toto}
\end{equation}
The first two terms represent the solvent contribution to the free
energy while $\Delta E_{0}$ is the difference in energy between the
2 solutes in vacuum. In this paper we restrict the study to rigid classical
solutes with no intramolecular potentials, so that last term vanishes.

In his original work, Marcus estimated the free energy cost to solvate
a solute within a solvent where polarization is not in equilibrium
with the electric field generated by the solute. In the MDFT framework
the density field contains all the  structural equilibrium information
of the solvent, including its polarization. We can consider  MDFT
 as a more general field theory than that used by Marcus.
Nevertheless, the density field itself remains a complicated object. To
facilitate our understanding it is useful to define a solvent reaction
coordinate \textit{i.e.} a scalar quantity which is uniquely defined
by the density field. 
By introducing a class of intermediate potentials interpolating between
state 0 and state 1, we show that the average vertical energy gap
is an appropriate order parameter. We then derive an expression for
the free energies of states 0 and 1 as  functions of the average vertical
energy gap.

States $0$ and $1$ are characterized by the following
Hamiltonian

\begin{equation}
H_{\eta}=K+U+V_{\eta}.\label{eq:Heta}
\end{equation}
 where $\eta=0$ or 1, $K$ is the kinetic energy and $U$
is the potential energy. For each state, the  equilibrium
probability distribution in the Grand Canonical ensemble is

\begin{equation}
f_{\eta}(\bm{X}^{N},\bm{p}^{N})=\Xi_{\eta}^{-1}\exp\left[-\beta\left(H_{\eta}\left(\bm{X}^{N},\bm{p}^{N}\right)-\mu N\right)\right],\label{eq:feta}
\end{equation}
where $\mu$ is the chemical potential of the solvent and $(\bm{X}^{N},\bm{p}^{N})$
is a point in  phase space $\bm{X}$ denotes 
the couple $\left(\bm{r},\bm{\Omega}\right)$ describing the CM 
and orientation of a solvent molecule with momentum $\bm{p}$. $\Xi_{\eta}$
is the corresponding grand partition function:

\begin{equation}
\Xi_{\eta}=\text{Tr}\left[\exp\left(-\beta\left(H_{\eta}-\mu N\right)\right)\right],\label{eq:Fparteta}
\end{equation}
where Tr denotes the classical trace
\begin{equation}
\text{Tr}\equiv\sum_{N=0}^{\infty}\frac{1}{h^{3N}N!}\int d\bm{X}_{1}...d\bm{X}_{N}\int d\bm{p}_{1}...d\bm{p}_{N}\label{eq:trace}
\end{equation}
and $h$ is the Planck constant. We now introduce a class of external
potentials defined as linear combinations of $V_{0}$ and $V_{1}$
\begin{equation}
V_{\eta}=V_{0}+\eta(V_{1}-V_{0})\text{ with }\eta\in\left[0,1\right].\label{eq:Veta-1-Appendix}
\end{equation}

This  defines the corresponding set of Hamiltonians (eq.\ref{eq:Heta}),
probability distributions (eq. \ref{eq:feta}) and grand partition
functions (eq.\ref{eq:Fparteta}) for any value of $\eta$. Since for physically relevant cases $V_{0}$ and $V_{1}$
differ by more than a constant, any value of $\eta$ defines a unique
potential $V_{\eta}$ (up to an irrelevant constant). Because of this
uniqueness of the potential, a unique equilibrium solvent density $\rho_{\eta}$ is associated with any value of $\eta$. This is a consequence of 
the cDFT principle \citep{evans_nature_1979} which implies a one-to-one
mapping between external potential, equilibrium distribution and equilibrium
solvent density \footnote[2]{Note that this is true for any $V_{\eta}=V_{0}+s(\eta)\left(V_{1}-V_{0}\right)$,
as long as $s$ is a strictly increasing continuous function with
$s(0)=0$ and $s(1)=1$.} .

We define the average vertical energy gap, related to an equilibrium
density $\rho_{\eta}$ by
\begin{equation}
\left\langle \Delta E\right\rangle _{\eta}=\iint\rho_{\eta}(\bm{r},\bm{\Omega})\left[V_{1}(\bm{r},\bm{\Omega})-V_{0}(\bm{r},\bm{\Omega})\right]d\bm{r}d\bm{\Omega}.\label{eq:DeltaEetaapp}
\end{equation}
This quantity represents the energy difference between states 1 and
0 solvated in the solvent of density $\rho_{\eta}$. We prove in Appendix \ref{App:equivalence_rhoeta_deltaEeta}
that, for the family of potentials in eq.\ref{eq:Veta-1-Appendix},
$\left\langle \Delta E\right\rangle _{\eta}$ is an adequate order parameter
since it uniquely defines $\rho_{\eta}$. Thus, the free energy of any state
is a function of $\left\langle \Delta E\right\rangle _{\eta}$. For instance, for state 0 it reduces to 
\begin{equation}
F_{0}\left(\left\langle \Delta E\right\rangle _{\eta}\right)\equiv F_{0}[\rho_{\eta}].\label{eq:F0=00005Beta=00005D_1}
\end{equation}

Note that the average free energy gap defined in eq.\ref{eq:DeltaEetaapp}
differs from the microscopic version, $\Delta E$,  used in MD;
\begin{equation}
\Delta E\left(\left\{ \bm{R}\right\} \right)=E_{1}\left(\left\{ \bm{R}\right\} \right)-E_{0}\left(\left\{ \bm{R}\right\} \right),\label{eq:DeltaE_MD-1}
\end{equation}
with $\left\{ \bm{R}\right\} $ denoting the whole set of coordinates
of solvent molecules, but they are actually related by
\begin{equation}
\left\langle \Delta E(\left\{ \bm{R}\right\} )\right\rangle _{\eta}=\left\langle \Delta E\right\rangle _{\eta},
\end{equation}
where on the left hand side $\left\langle ..\right\rangle _{\eta}$
denotes the thermodynamic average on the potential energy surface
$\eta$. $\left\langle \Delta E\right\rangle _{\eta}$ is also frequently
 reported in MD studies of ET since it is another measure of the validity
of Marcus Theory which predicts that it varies linearly with the coupling
parameter. Our approach is  closer to Marcus' original work
\citep{marcus_exchange_1960} where he mentioned that the ``equivalent
equilibrium distribution would be obtained in a corresponding equilibrium
system in which the charges on the two central ions'' are linear
combinations of the original ones.

We can now express the reorganization free energies displayed in Fig.\ \ref{fig:Marcus_Plot} as

\begin{equation}
\lambda_{0}=F_{0}\left(\left\langle \Delta E\right\rangle _{1}\right)-F_{0}\left(\left\langle \Delta E\right\rangle _{0}\right)=F_{0}[\rho_{1}]-F_{0}[\rho_{0}]=\Delta W-\left\langle \Delta E\right\rangle _{1},\label{eq:lamda_a}
\end{equation}
\begin{equation}
\lambda_{1}=F_{1}\left(\left\langle \Delta E\right\rangle _{0}\right)-F_{1}\left(\left\langle \Delta E\right\rangle _{1}\right)=F_{1}[\rho_{0}]-F_{1}[\rho_{1}]=-\Delta W+\left\langle \Delta E\right\rangle _{2}.\label{eq:lamda_b}
\end{equation}
Borgis and coworkers have  reported a similar relation to compute
the reorganization free energies using MDFT \citep{borgis_molecular_2012}.
Under the assumption that Marcus theory is valid - hence that the
two reorganizations free energies are equal, \textit{i.e.} $\lambda=\lambda_{0}=\lambda_{1}$,  
eq. \ref{eq:lamda_a} and eq. \ref{eq:lamda_b} reduce to:
\begin{align}
\lambda & =  \frac{\lambda_{0}+\lambda_{1}}{2}=\frac{\left(F_{0}[\rho_{1}]-F_{1}[\rho_{1}]\right)-\left(F_{0}[\rho_{0}]-F_{1}[\rho_{0}]\right)}{2}\nonumber \\
 & =  \frac{1}{2}\iint\left[V_{1}(\bm{r},\bm{\Omega})-V_{0}(\bm{r},\bm{\Omega})\right]\left[\rho_{0}(\bm{r},\bm{\Omega})-\rho_{1}(\bm{r},\bm{\Omega})\right]d\bm{r}d\bm{\Omega}\label{eq:.dkvk.}
\end{align}
which is equivalent to the linear response formula ${\lambda=\frac{1}{2}\left(\left\langle \Delta E\right\rangle _{0}-\left\langle \Delta E\right\rangle _{1}\right)}$
often used in molecular simulations. We also note that, as for the
usual variable $\Delta E$ in MD, the exact relation introduced
by Warshel is satisfied:
\begin{equation}
F_{1}\left(\left\langle \Delta E\right\rangle _{\eta}\right)=F_{0}\left(\left\langle \Delta E\right\rangle _{\eta}\right)+\left\langle \Delta E\right\rangle _{\eta}.
\end{equation}
This is a corollary of eq.\ref{eq:Legendre} with $\eta=1$. In the
next subsection we explain how the average vertical energy gap, reorganization
free energies and free energy curves are computed using MDFT.

\subsubsection{Computational details}

To study a given system, we shall minimize functionals corresponding
to different external potentials $V_{\eta}$ according to eq.\ref{eq:Veta-1-Appendix}.
We  consider only cases for which the Lennard-Jones sites of the solute
remain unchanged between state $0$ and state $1$, so that the energy
gap reduces to the difference in the electrostatic potential energy
of the solute in the field generated by the solvent molecules. This
can be computed using the electrostatic potential generated by states
$0$ and $1$, while the vertical energy gap can  be computed using
eq.\ref{eq:DeltaEetaapp}.

As shown above, the free energies corresponding to these values of
the energy gaps are
\begin{equation}
F_{A}\left(\left\langle \Delta E\right\rangle _{\eta}\right)=F_{A}\left[\rho_{\eta}\right]\label{eq:DeltaF direct way}
\end{equation}
with $A=0,1$. To construct the FEC as in Fig.\ \ref{fig:Marcus_Plot}
we first minimize the functional of eq.\ref{eq:F=00003DFid+Fexc+Fext}
for several values of $\eta$ to obtain $\rho_{\eta}$, next compute
the value of the average vertical energy gap, and finally evaluate $F_{0}$
and $F_{1}$ for the different $\rho_{\eta}$.

An alternative route to compute the FEC was previously proposed
by Hirata \textit{et al.} \citep{chong_free_1996,sato_theoretical_2003}
using another implicit solvent method, RISM. We show in Appendix \ref{sec:Appendix-B:-Thermodynamic}
that the thermodynamic cycle they propose is equivalent to the present
scheme, although not expressed in a free energy density functional
language. Now that we have shown how MDFT can be used to investigate
ET reactions, the following  section is dedicated to assesses the validity
of this approach on simple and  complex solutes.

\section{Applications\label{sec:Implementation}}

\subsection{ET between Cl$^{0}$, Cl$^{+}$ and Cl$^{-}$ ions\label{subsec:ET-between-Cl,}}

In this article we focus on the difficult case of aqueous solvation,
but calculation for simpler solvents such as acetonitrile or CO$_{2}$
are expected to give results of comparable  quality. We apply
the necessary correction due to periodic boundary conditions to charged
solutes \citep{kastenholz_computation_2006,kastenholz_computation_2006-1}
and an additional correction accounting for the  overestimation of the pressure
within HNC \citep{jeanmairet_molecular_2015} to both neutral and charged solutes.

To allow comparisons, we chose a system which has been extensively
studied using MD by Hartnig \textit{et al.} \citep{hartnig_molecular_2001}.
This model of chlorine consists of one Lennard-Jones site, with $\sigma=4.404 \ \textrm{\AA}$
and $\epsilon=$ 0.4190 kJ.mol$^{-1}$, and a charge equal to ${-1}$,
0 or 1  elementary charges $e$. To compute the FEC of the atom
and the 2 ions with a good accuracy we ran MDFT calculations with
a solute charge varying in steps of 0.1 elementary charges.  We used
a $40\times40\times40 \ \textrm{\AA\ensuremath{^{3}}}$ box with 120$^{3}$
spatial grid points and 196 possible orientations per spatial point.
The solvent is SPC/E water for which the exact direct correlation
function projected on a basis of rotational invariants was obtained
by Belloni\textit{ et al.} using a hybrid Monte Carlo plus Integral
Equation approach \citep{puibasset_bridge_2012,belloni_efficient_2014,belloni_exact_2017}.
All  simulations are carried out at 298.15 K.

\begin{figure}
\begin{centering}
\begin{tabular}{c}
\includegraphics[width=0.4\textwidth]{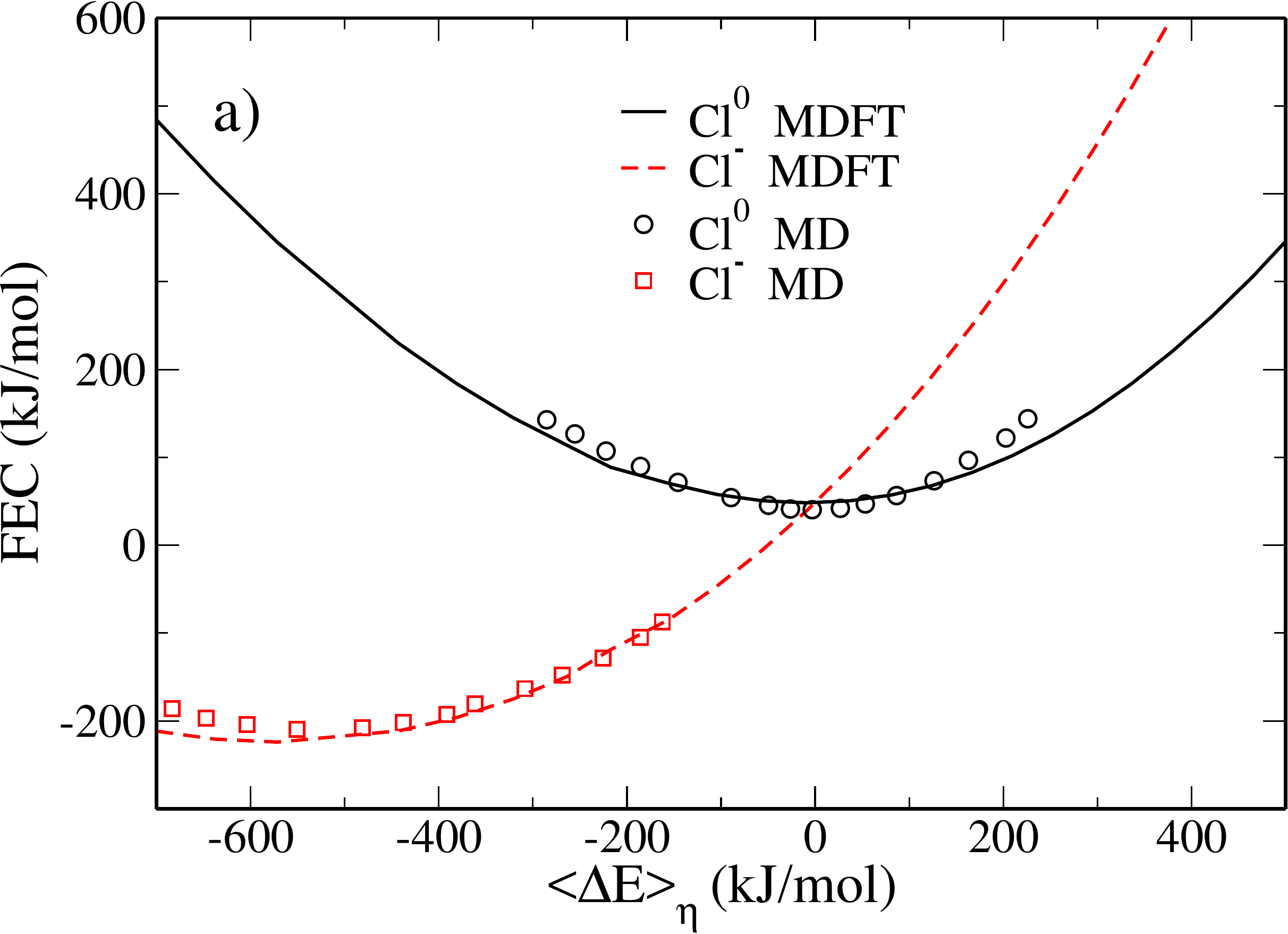}\tabularnewline
\includegraphics[width=0.4\textwidth]{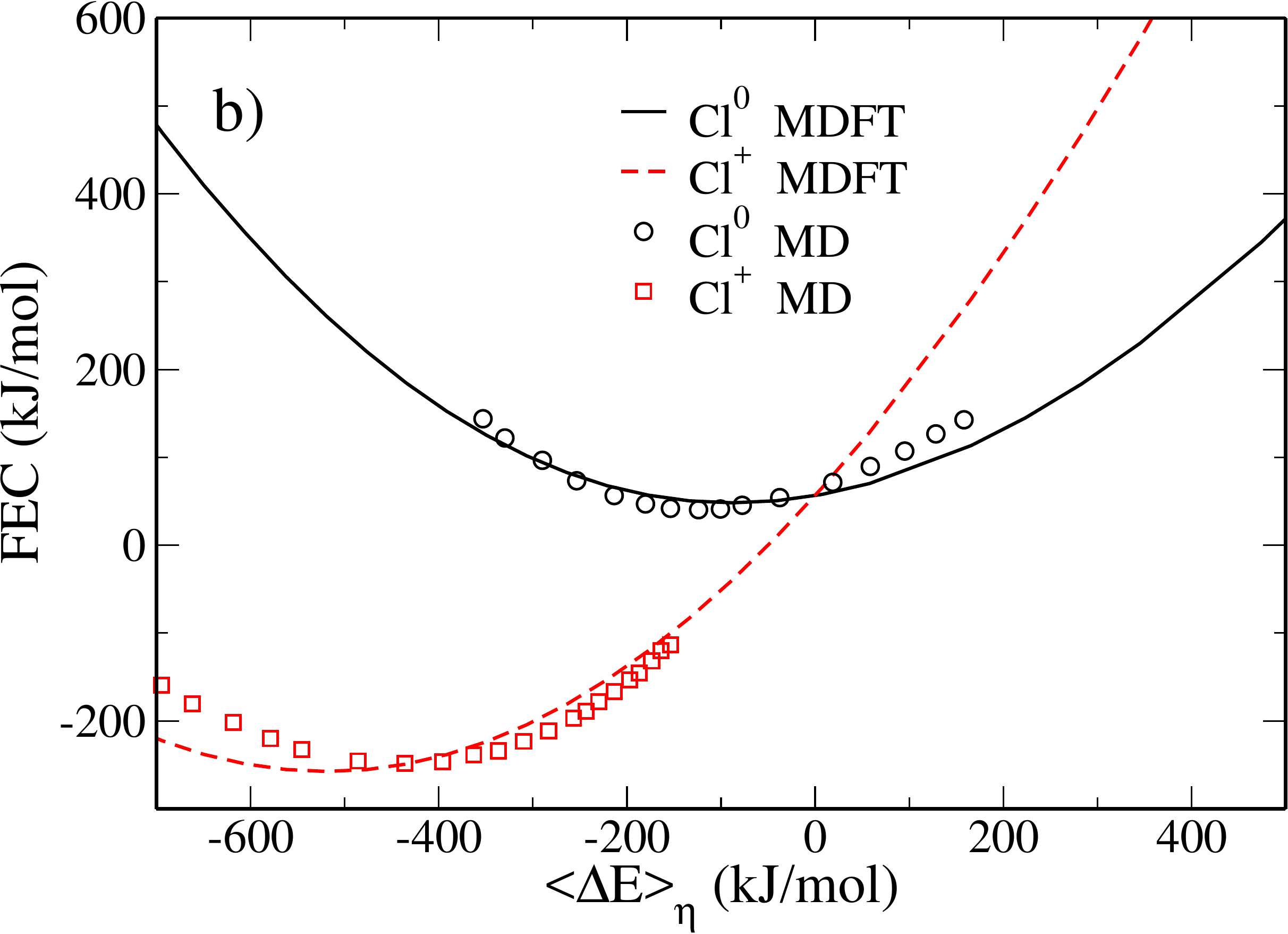}
\end{tabular}
\end{centering}
\caption{Pairs of free energy curves of a) Cl$^{0}$/Cl$^{-}$ and b) Cl$^{0}$/Cl$^{+}$
 as a function of the average vertical energy gap. The black solid
line and the dashed red line correspond to the MDFT results for the
atom and the ions, respectively. Those results are compared to  Hartnig's work \citep{hartnig_molecular_2001}
reported as a function of the absolute vertical energy gap. The black
circles correspond to the atom and the red squares to the ions.\label{fig:Wcl}}
\end{figure}

The FEC are shown in Fig.\ \ref{fig:Wcl} which compares the MDFT
results (solid lines) to the MD results \citep{hartnig_molecular_2001}
(symbols). The representation adopted here differs from that used
by Hartnig and Koper, since we report the FEC as a function of the
vertical energy gap and do not apply an arbitrary vertical shift of
the curves. The methodology used to plot the MD data in this representation
is described in Appendix \ref{sec:Appendix-C:-Modification}. This
representation is better suited to highlight some features of the
ET. For instance, we note that both pairs of FEC cross
when the average free energy gap is equal to 0, as expected. 

The agreement between MD and MDFT is satisfactory, the main difference being
a shift of the MDFT curves towards  negative values of $\left\langle \Delta E\right\rangle _{\eta}$
for the cation. The most interesting observation from Fig.\ \ref{fig:Wcl}
is the consistency between the curvature of the curves obtained by
MD and by MDFT. In particular, Cl$^{0}$ and Cl$^{+}$ exhibit a similar
curvature while that of Cl$^{-}$ is larger, indicating that the neutral
to anion ET does not follow the Marcus picture while the neutral to
cation ET does. To be more quantitative, we computed the reorganization
free energy associated with the three species, based on the curvature
by fitting the data within 90 kJ/mol from the minimum with the following
expression
\begin{equation}
F_{A}[\Delta E]=\frac{1}{4\lambda}\left(\Delta E-\Delta E_{min}\right)^{2},\label{eq:fitlamfa}
\end{equation}
 a strategy that was adopted by Hartnig \textit{et al}. This
assumes that the curve can be fitted by a parabola. The expression
linking the curvature and the parabola parameter in eq.\ref{eq:fitlamfa}
is derived using Marcus theory.

The reorganization free energies obtained via MDFT and MD are compared
in Table \ref{tab:lamdaCl}. For the MD data, we report the values
of the free energies given in the original work, in addition to the
one we have recomputed to keep the fitting procedure consistent between
the two approaches. MDFT overestimates the reorganization free energies 
compared to MD. However, comparing the values of the reorganization
free energies between species leads to conclusions similar to those concerning the curvature of the FEC. The neutral atom and the
cation have a similar reorganization free energy, while the anion
has a much larger one. 

\begin{table}
\small
   \caption{ Reorganization free energies computed via eq.\ref{eq:fitlamfa} using
data below 90 kJ/mol from Fig.\ \ref{fig:Wcl}. For MD, we recomputed
the reorganization free energy using points extracted from figure 4 of
Ref. \cite{hartnig_molecular_2001}. In parenthesis we report the
original data of that publication}
\label{tab:lamdaCl}
   \begin{tabular*}{0.48\textwidth}{@{\extracolsep{\fill}}lll}
    \hline
    Species & $\lambda_{\text{MDFT}}$ (kJ/mol) &$\lambda_{\text{MD}}$ (kJ/mol)\tabularnewline
    \hline
Cl$^{0}$ & 233 & 153 (132)\tabularnewline
Cl$^{-}$ & 297 & 263 (252)\tabularnewline
Cl$^{+}$ & 216 & 165 (177)\tabularnewline
    \hline
  \end{tabular*}
\end{table}

The simple picture emerging from the comparison of curvatures
is however misleading. Table \ref{tab:lamdaCl}  suggests that a single
 reorganization free energy can be associated with each solute, but this
 does not hold for several reasons: i) it assumes that the FEC of a solute
is a  parabola ii) it neglects the other solutes involved in the
ET. We should refer to the reorganization free energy for a given $0\rightarrow1$
ET reaction as defined in Fig.\ \ref{fig:Marcus_Plot}, because the
meaningful physical quantity is a free energy, not the curvature of
a fitting curve. To illustrate this point, we report in Table \ref{tab:lamdaCl-2_DeltaF}
the reorganization free energies computed using eq.\ref{eq:lamda_a}
and eq.\ref{eq:lamda_b}. These  free energies for the
Cl$^{0}\rightarrow$ Cl$^{+}$ reaction are almost identical  
to those reported in Table \ref{tab:lamdaCl}. This is consistent
with the Marcus picture: If the two FEC are identical parabolas, there
is a unique $\lambda$ parameter defining the curvature and the two
free energy differences. However, in the Cl$^{0}\rightarrow$ Cl$^{-}$
ET reaction the reorganization free energy of state 0 is much larger
than in the other ET reaction. This is a consequence of the larger
curvature of the second state. The reorganization free energy of Cl$^{-}$
is significantly reduced compared to those listed in Table \ref{tab:lamdaCl}, another
consequence of the smaller curvature of state 0. 

\begin{table}
\small
   \caption{Reorganization free energies computed by MDFT via eq.\ref{eq:lamda_a} and eq.\ref{eq:lamda_b} for both ET reactions}
	\label{tab:lamdaCl-2_DeltaF}
   \begin{tabular*}{0.48\textwidth}{@{\extracolsep{\fill}}lll}
    \hline
    Species & Cl$^{0}/$Cl$^{-}$ (kJ/mol) & Cl$^{0}$/Cl$^{+}$ (kJ/mol)\tabularnewline
    \hline
   Cl$^{0}$ & 297 & 214\tabularnewline
	Cl$^{-}$ & 264 & N/A\tabularnewline
	Cl$^{+}$ & N/A & 218\tabularnewline
    \hline
  \end{tabular*}
\end{table}

This difference between the reorganization free energies is a further
indication that the 0 to -1 ET does not follow Marcus theory. 

\begin{figure}
\begin{centering}
\includegraphics[width=0.4\textwidth]{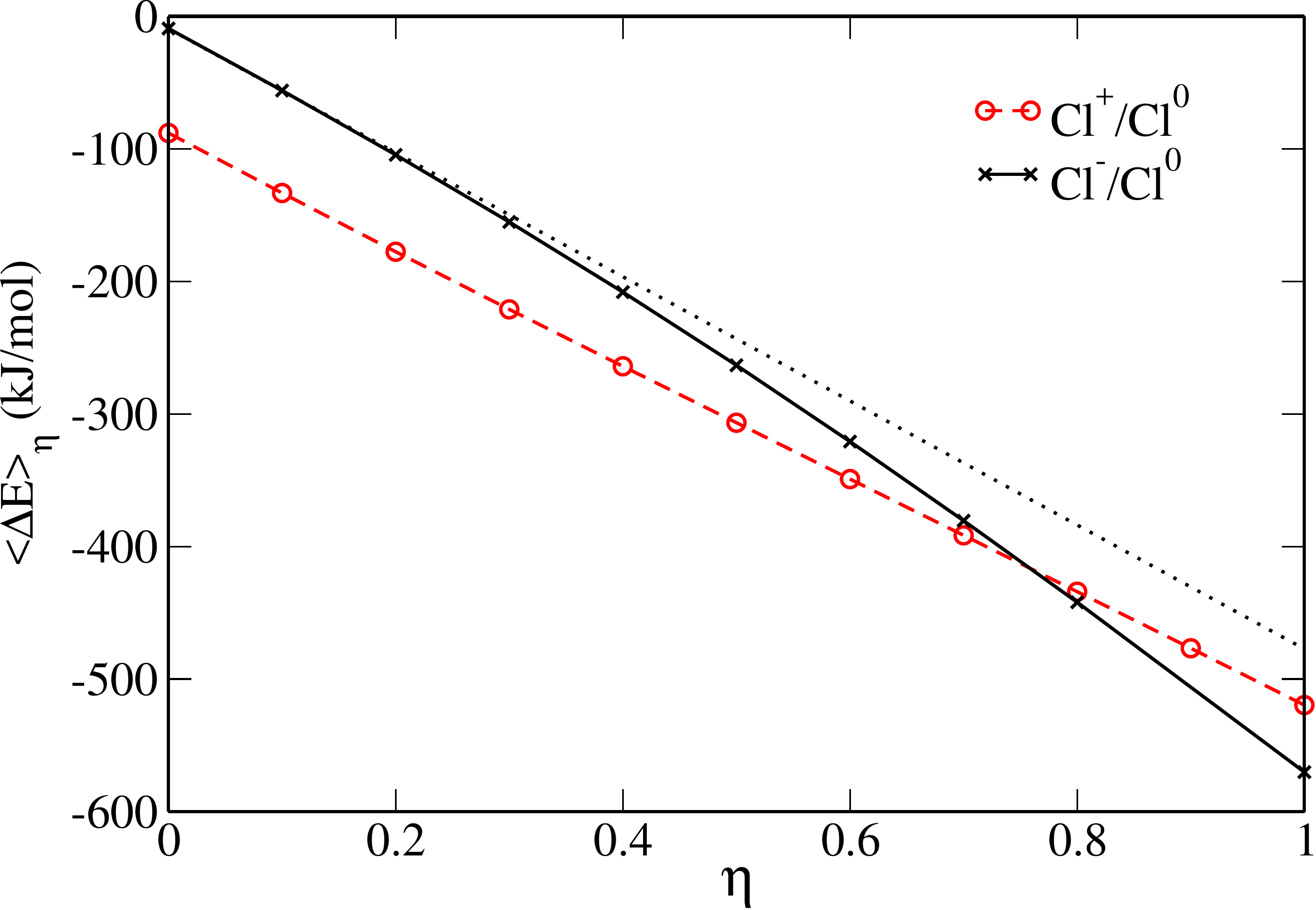}
\par\end{centering}
\caption{Average value of the vertical energy gap versus the coupling
parameter $\eta$. The Cl$^{0}\rightarrow$ Cl$^{-}$ ET reaction
is shown in solid black, the Cl$^{0}\rightarrow$ Cl$^{+}$ in dashed red.
The dotted curve is a linear fit to the first values of vertical energy
gap in the case of the ET reaction involving the anion, it is shown as a guide to the eye.\label{fig:DeltaEeta}}
\end{figure}

Another way to check if the ET reaction follows the Marcus picture
is to consider the evolution of the average vertical energy gap $\left\langle \Delta E\right\rangle _{\eta}$
with the parameter $\eta$. As mentioned earlier, such a curve is linear
in Marcus theory. The evolution of the average vertical energy gap
with $\eta$ is presented in Fig.\ \ref{fig:DeltaEeta}.
For the neutral to positive charge transfer, the vertical energy gap
does vary linearly with the coupling parameter $\eta$. In contrast,
for the neutral to anion ET a non-linear variation is observed indicating
a deviation from Marcus theory. 

Thus, the curvatures of the FEC, the
values of the reorganization free energies and the variation of $\left\langle \Delta E\right\rangle _{\eta}$
with respect to $\eta$ consistently indicate a different behavior
for the two ET reactions. This has already been noticed by Hartnig\textit{
et al,} who rationalized this observation by arguing that while the
distance between the solute and the oxygen of the first solvation
layer remains similar for all  oxidation numbers, the hydrogen
is much closer to the solute in the case of the anion. This causes
a ``shrinking'' of the first solvation shell in the case of Cl$^{-}$
which differs considerably from the solvation shell of the neutral and positive
solutes. Such a difference in the solvation shells of the two species
cannot be properly captured by linear response assumed in Marcus
Theory. 

\begin{figure}
\begin{centering}
\includegraphics[width=0.4\textwidth]{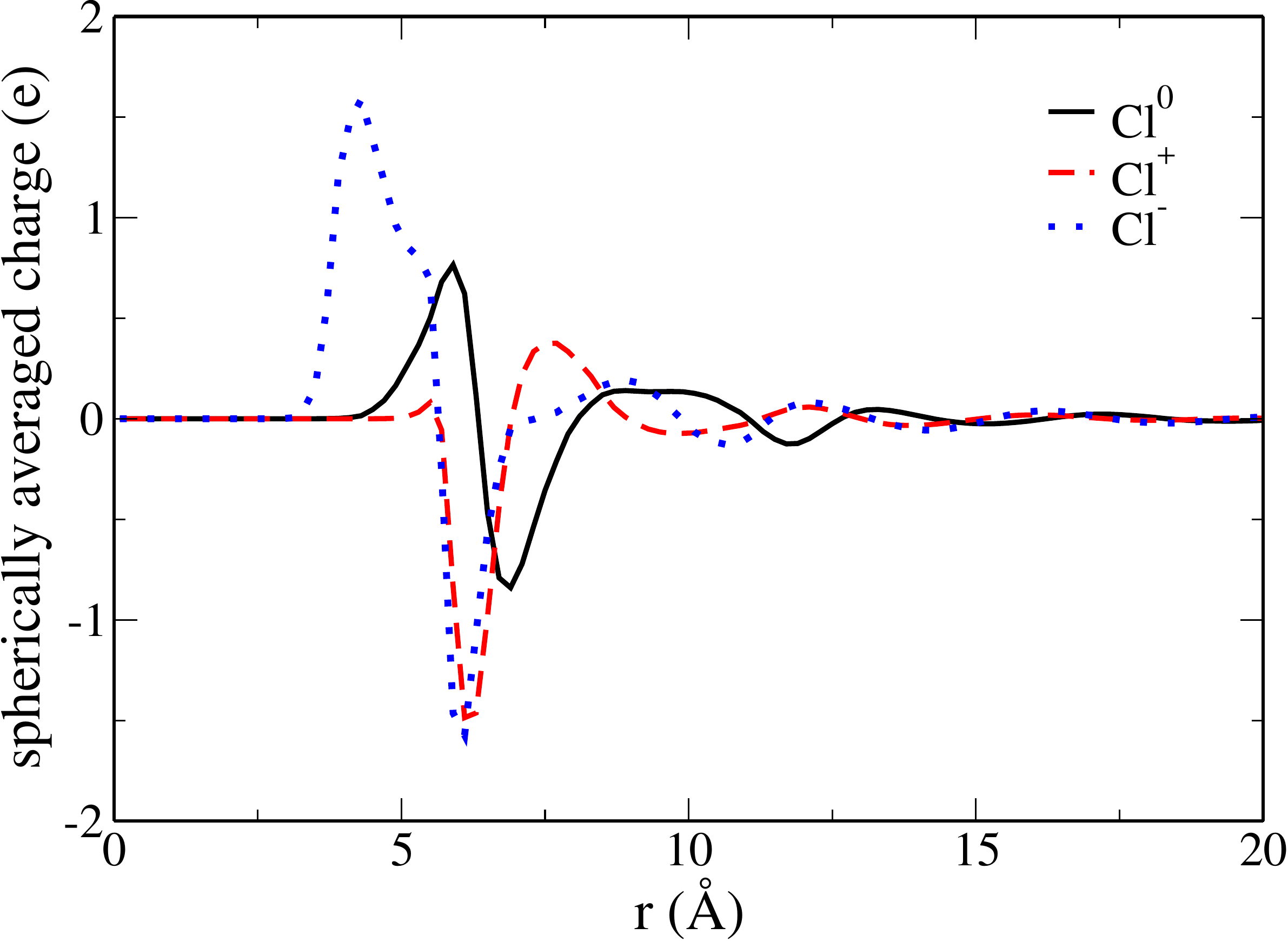}\tabularnewline
\par\end{centering}
\caption{Spherically averaged solvent charge density as a function of the distance to the solute. The curve corresponding to the neutral solute is in full black, the one of the cation is in dashed red and the one of the anion in dotted blue.
 \label{fig:Clstrucutr}}
\end{figure}

Since MDFT gives access to the solvent density, we can also investigate
the solvation structure. We compute the solvent charge density 

\begin{equation}
\rho^{c}(\bm{r})=\iint\rho(\bm{r}^{\prime},\bm{\Omega})\sigma(\bm{r}-\bm{r}^{\prime},\bm{\Omega})d\bm{r}^{\prime}d\bm{\Omega},\label{eq:rho_c-2}
\end{equation}
where $\sigma(\bm{r},\bm{\Omega})$ is the charge distribution at
point $\bm{r}$ of a single solvent molecule located at the origin,
with the orientation $\bm{\Omega}$
\begin{equation}
\sigma(\bm{r},\bm{\Omega})=\sum_{i}q_{i}\delta(\bm{r}-\bm{r}_{i\bm{\Omega}})
\end{equation}
where the sum runs over the solvent sites, $\delta$ is the Dirac
distribution, $\bm{r}_{i\bm{\Omega}}$ is the position of site $i$
and $q_{i}$ its charge. The spherically averaged one-dimensional solvent charge densities are reported in Fig.\ \ref{fig:Clstrucutr}
for the 3 oxidation numbers as a function of the distance to the solute. For all solutes, we observe a zone
of zero charge density for small values of $r$, \textit{i.e.} close to the solute, corresponding to the absence
of water molecules. Then, alternating regions of positive and negative
charge point to a preferential orientation of the solvent in the
solvation shells. Finally,  zero charge density is reached far from
the solute at large $r$ when a bulk behavior without preferential orientation is recovered. 

If we first consider the neutral and positive solutes, we observe
in Fig.\ \ref{fig:Clstrucutr} that the preferential orientation of water
in the first solvation shell reverses between the neutral solute and
the cation. Around the cation, the water molecules in the first solvation
shell have their oxygen pointing toward the solute. For the neutral
species the hydrogen of water molecules are the closest
to the solute. However the positions of the first extrema are similar, 5.9 $\textrm{\AA}$ for Cl$^0$, 6.3 $\textrm{\AA}$ for Cl$^+$. This indicates
that the two solvation shells essentially differ by the orientation
of the water molecules. On the contrary, solvent molecules are much closer to the anion, where the first maximum originating from hydrogen is located at 4.3 $\textrm{\AA}$. There is a
shrinking of the first solvation shell for the anion, in agreement
with Hartnig \textit{et al. }\citep{hartnig_molecular_2001}. It is
confirmed by the comparison of the partial molar volume computed thanks
to the equilibrium densities, namely 60 $\textrm{\AA}^{3}$ for Cl$^{0}$ and
Cl$^{+}$ and 6 $\textrm{\AA}^{3}$ for Cl$^{-}$. This difference in
the solvation shell explains why Marcus theory fails to
describe this ET reaction.

\subsection{Solid/solvent interface}

We now turn to the study of the influence of a solid/solvent interface on the ET
reaction. There are only few such studies available due to the computational
cost of MD which is to date the only simulation tool used in
this context.  It is worth mentioning the investigations by Remsing \textit{et al.} \citep{remsing_frustrated_2015}
who used MD and   by Li \textit{et al.} \citep{li_confinement_2017}
based on coarse grained MD. In the former, ions are highly confined between
two MnO$_{2}$ sheets and confinement is kept constant throughout the study. In the supplementary material of Li's article, the
authors report the evolution of the reorganization free energy when
the ion moves towards graphite sheets. They used umbrella sampling
to constrain the position of the redox active site in the direction
$z$ perpendicular to the surface but no constraint was applied on
the lateral coordinates. The dependence of the reorganization free-energy
on the distance between solute and electrode was subsequently obtained
through binning in the $z$-direction. In this set-up the position
of the solute is not frozen but may fluctuate around 
value of $z$ under consideration, taking all possible values in $x$ and $y$.
The reported reorganization free energy is hence a statistical average.

The computational efficiency of MDFT allows a systematic study of
the evolution of the reorganization free energy when the solute carrying 
the charge approaches an atomistically resolved wall. Because the
solute is kept fixed in the MDFT calculation it is not necessary to
resort to biasing techniques to constrain its position and no fluctuations
blur the reorganization free energies. We consider the Cl$^{0}\rightarrow \ $Cl$^{+}$
ET with the parameters introduced in section \ref{subsec:ET-between-Cl,}
and study the influence of the proximity of a wall made up 
of 400 atoms arranged as the (100) surface of a fcc crystal. The size of the wall
is $40\times40\ \textrm{\AA}^{2}$and the distance between neighbouring 
atoms is 2 $\textrm{\AA}$. Each atom is modeled by a Lennard-Jones
site with parameters $\sigma=3.37\ \textrm{\AA}$ and $\epsilon=0.23$
kJ/mol similar to that used to model graphite atoms in previous
studies \citep{cole_interaction_1983}. To study separately the effect
of the solvent on the ET we remove direct interactions between the
solute and the wall. We used a $40\times40\times40\ \textrm{\AA}^{3}$
cubic box with 3 grid points per $\textrm{\AA}$ and 196 discrete orientations
per grid point.

\begin{figure}
\centering{}\includegraphics[width=0.4\textwidth]{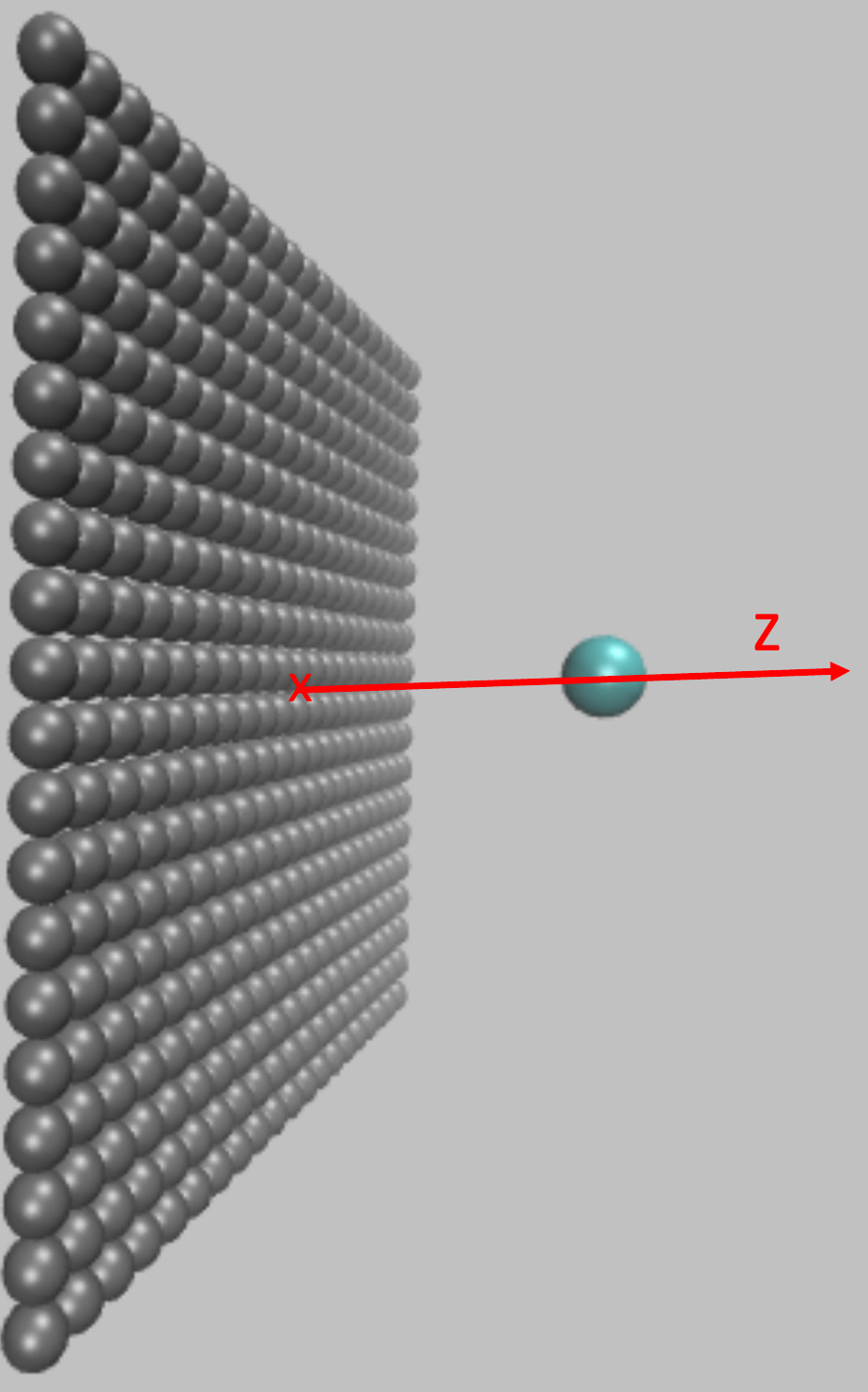}\caption{Snapshot of   system under consideration: the flat wall is shown in grey,
the solute in blue. The solute is moved along the $z$ direction perpendicular
to the wall\label{FIG:WALL}.}
\end{figure}

\begin{figure}
\centering{}%
\begin{tabular}{c}
\hspace{-0.1cm} \includegraphics[width=0.4\textwidth]{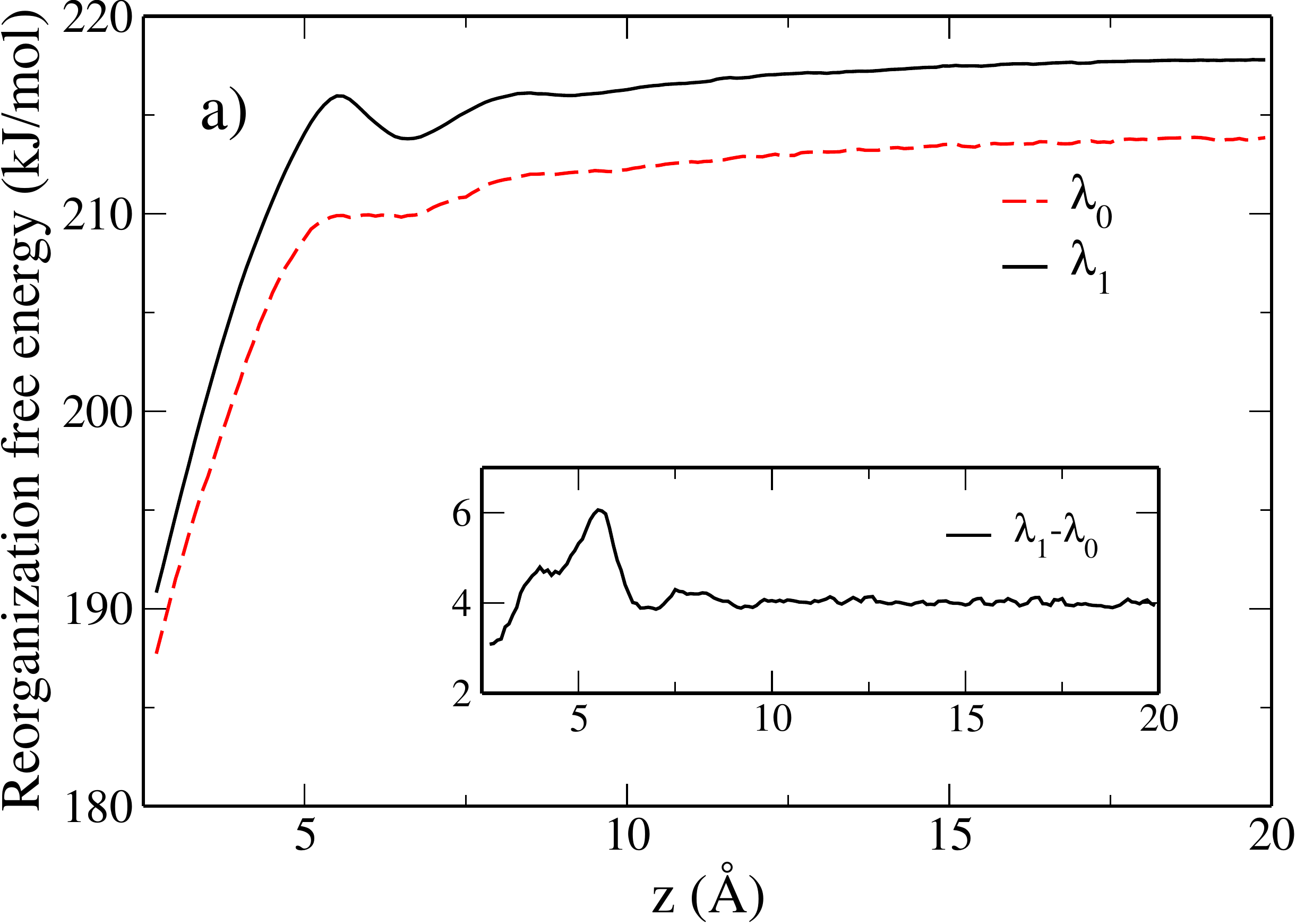}            \tabularnewline
\hspace{-0.2cm} \includegraphics[width=0.4\textwidth]{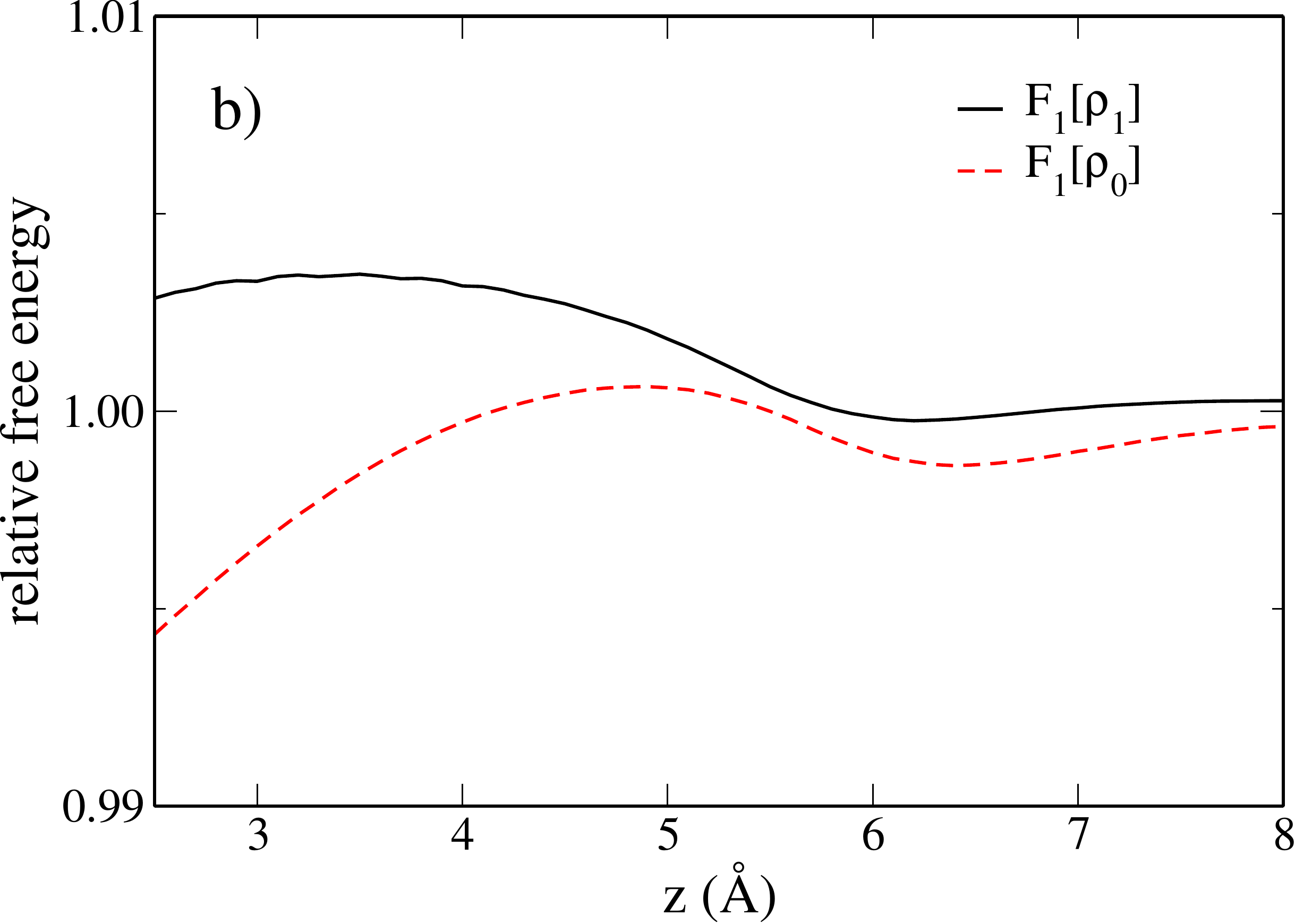}\tabularnewline
 \includegraphics[width=0.4\textwidth]{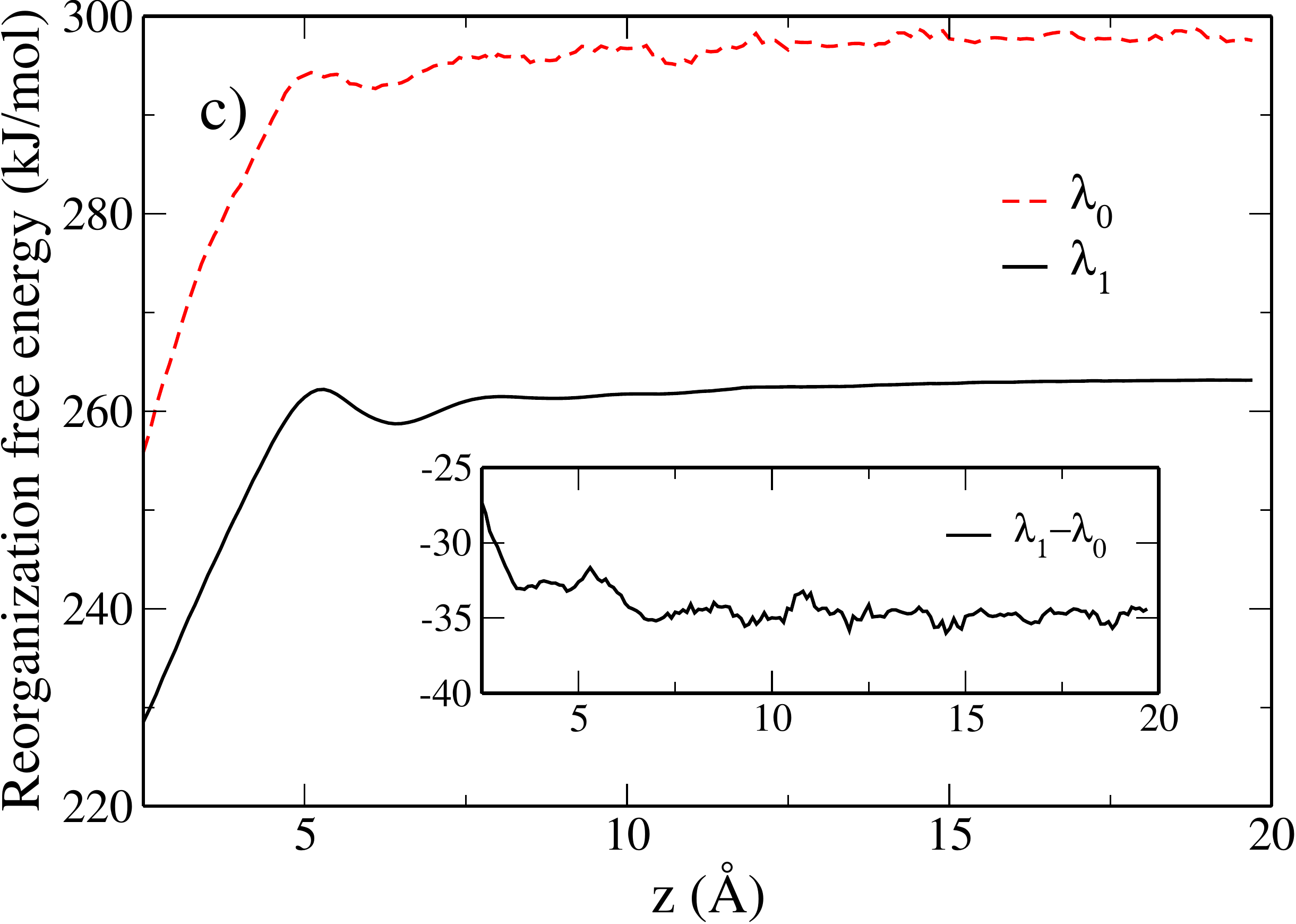}\tabularnewline
\end{tabular}\caption{a) Variation of the reorganization free energy for the Cl$^0 \ \rightarrow$ Cl$^+$ ET with the  
distance between the solute and the wall. The curve corresponding to the 
neutral state is displayed in dashed red, the one for the charged
solute in dotted black. The difference between $\lambda_1$ and $\lambda_0$ is shown in the inset. b)
zoom on the two components $F_{1}[\rho_{0}]$ and $F_{1}[\rho_{1}]$
of $\lambda_{1}$ for the Cl$^0 \ \rightarrow$ Cl$^+$ ET, the two quantities have been normalized by their
bulk value to assist visualization. c) Same as plot a) for the  Cl$^0 \ \rightarrow$ Cl$^-$ ET \label{FIG:lamda_wall}}
\end{figure}

We move the solute along the $z$ axis perpendicular to the surface
as illustrated in Fig.\ \ref{FIG:WALL}, with 175 calculations from
$z=2.5\ \textrm{\AA}$ to $z$ = 20 $\textrm{\AA}$ in steps of $dz=0.1\ \textrm{\AA}$.
The reorganization free energies of  the Cl$^0\ \rightarrow$ Cl$^+$ ET computed using eq.\ref{eq:lamda_a}
are displayed in panel a) of Fig.\ \ref{FIG:lamda_wall} in dotted
black for the charged solute and in dashed red for the neutral solute.
The two curves are similar and differ by less than 3 kJ/mol. This
is a small difference consistent with the result of Li et al. \citep{li_confinement_2017}
who reported that the ET of an iron atom dissolved in an ionic liquid
next to a polarizable planar electrode follows Marcus' scenario.

\begin{figure*}[]
\centering{}%
\begin{tabular}{cc}
\includegraphics[width=0.2\textwidth]{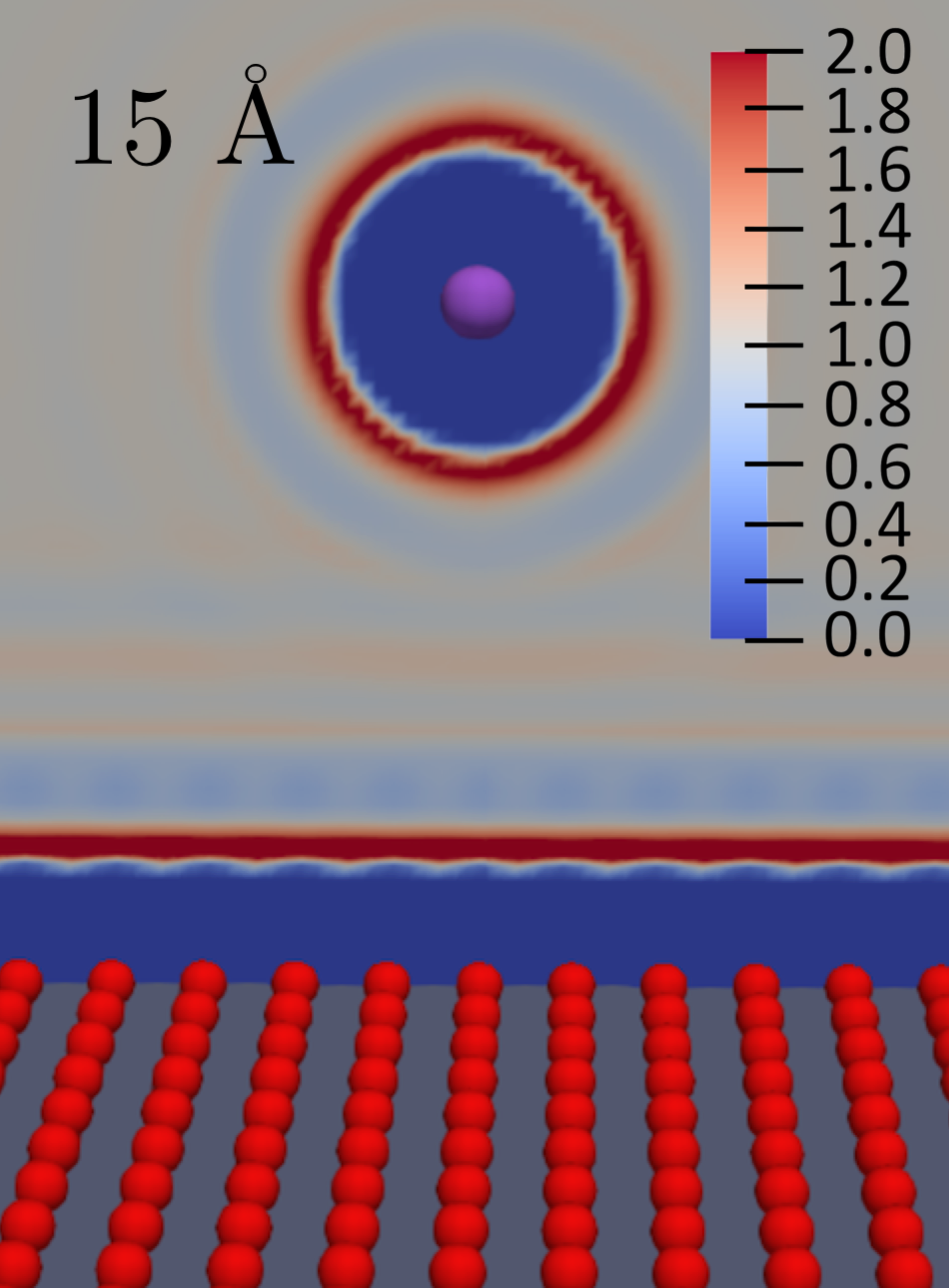} & \includegraphics[width=0.2\textwidth]{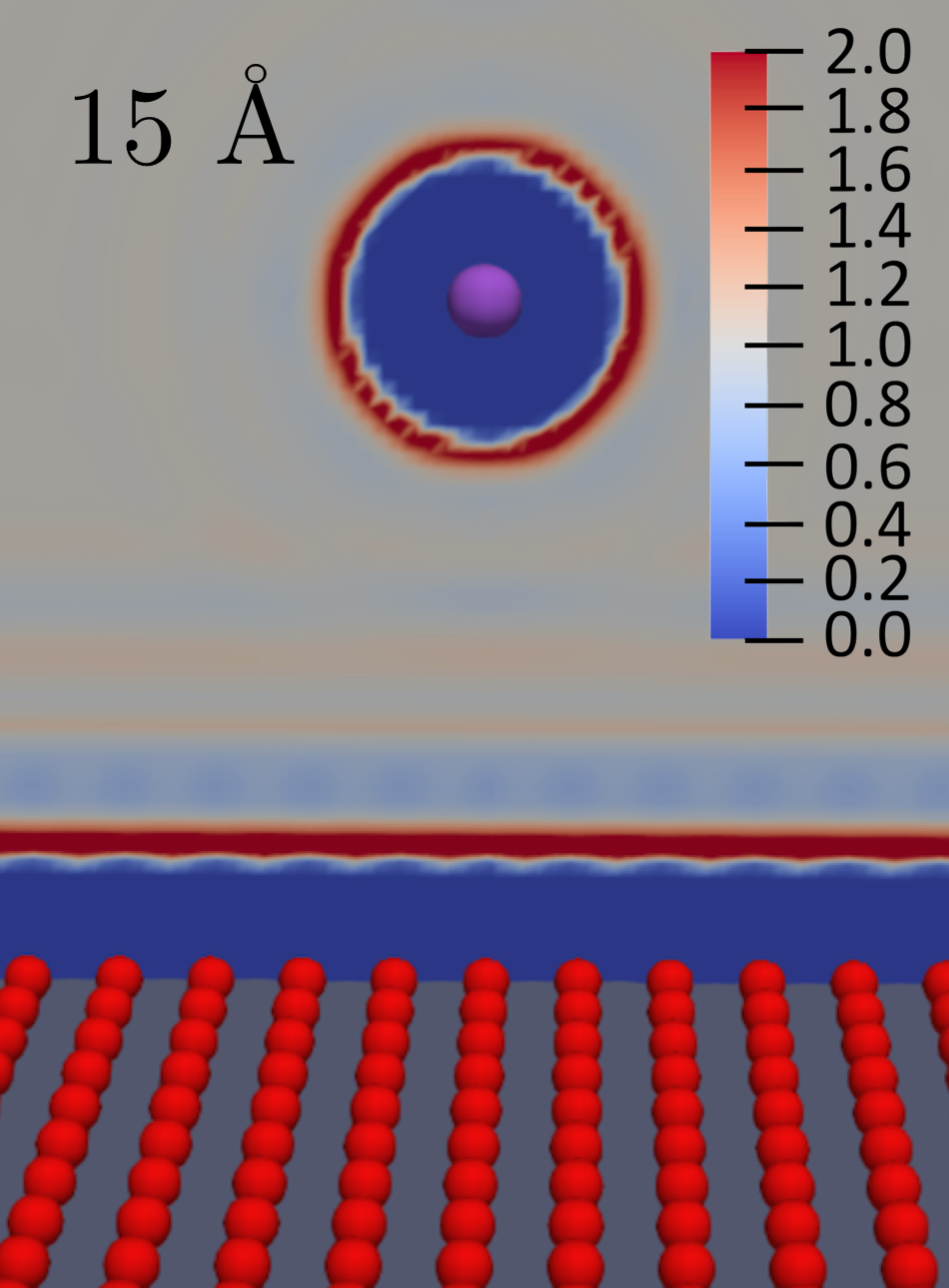}\tabularnewline
\includegraphics[width=0.2\textwidth]{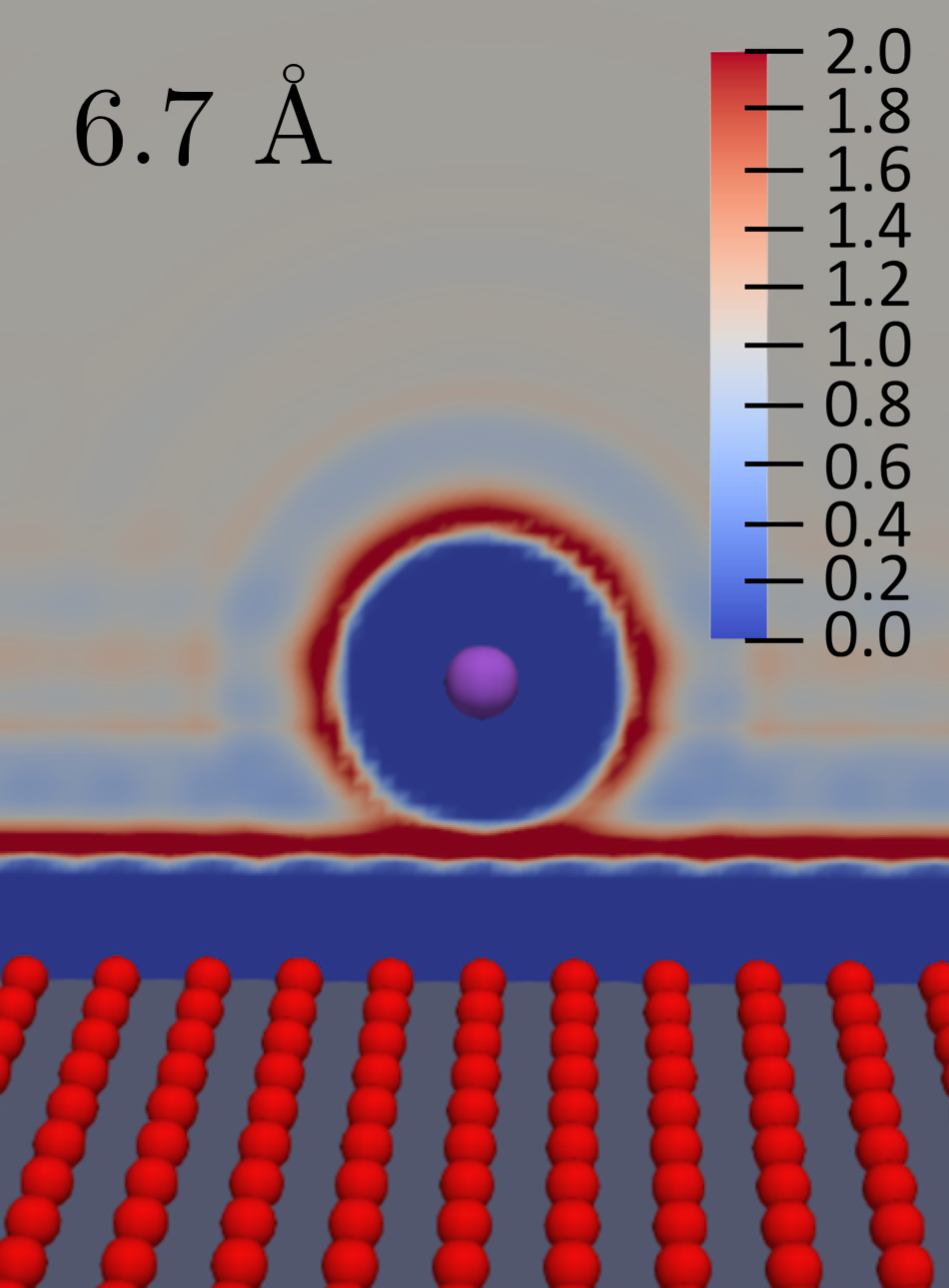} & \includegraphics[width=0.2\textwidth]{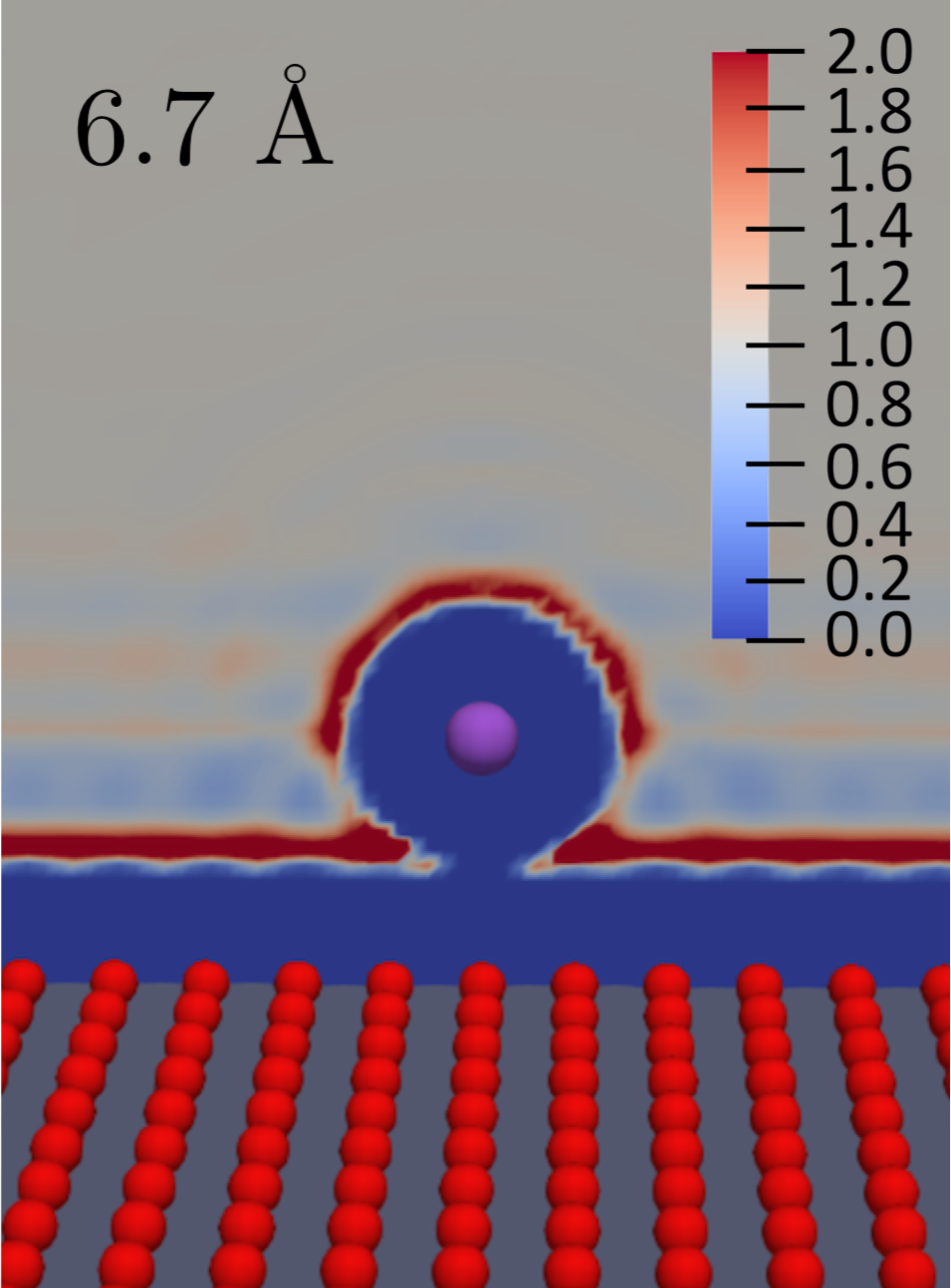}\tabularnewline
\includegraphics[width=0.2\textwidth]{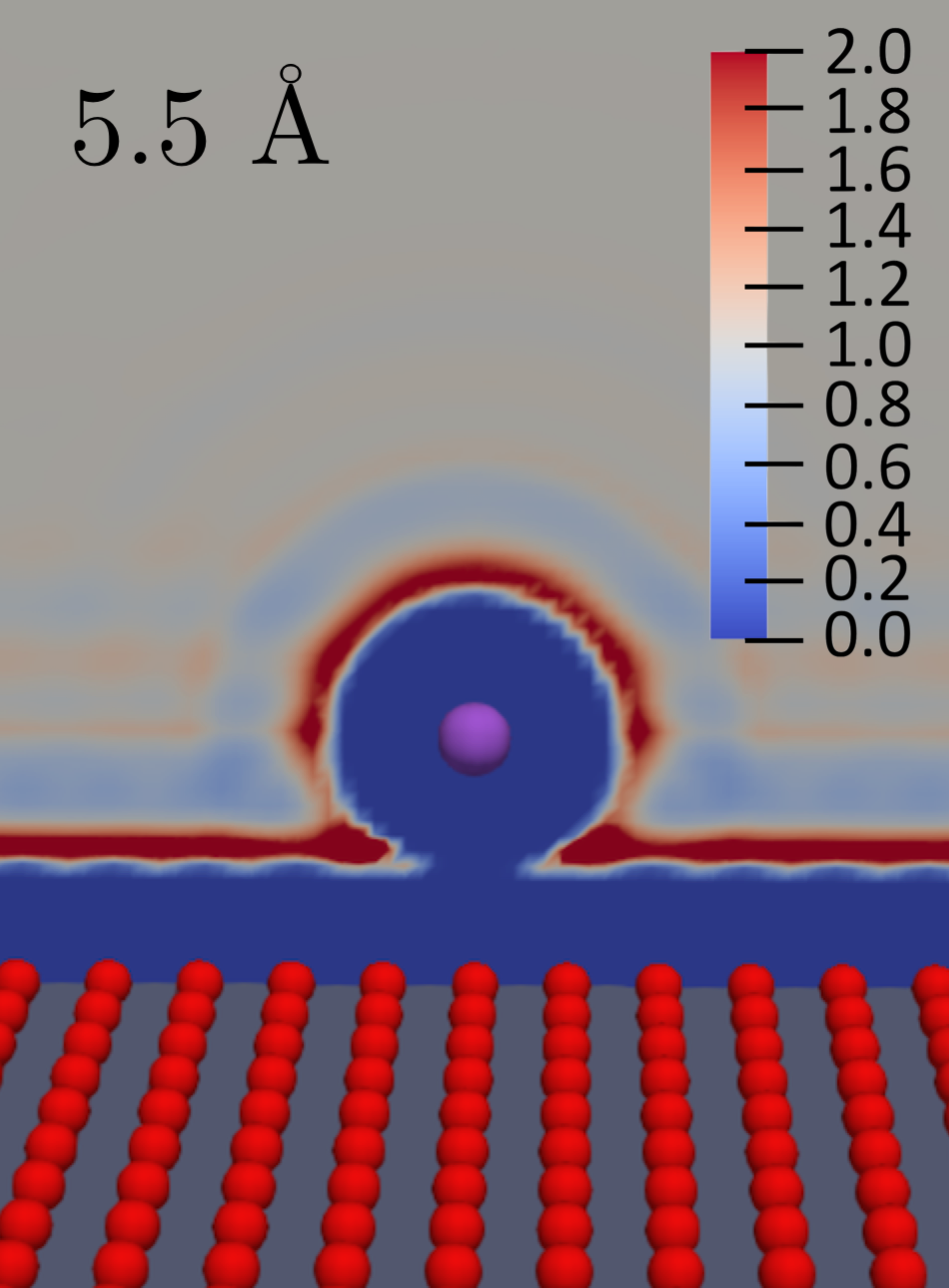} & \includegraphics[width=0.2\textwidth]{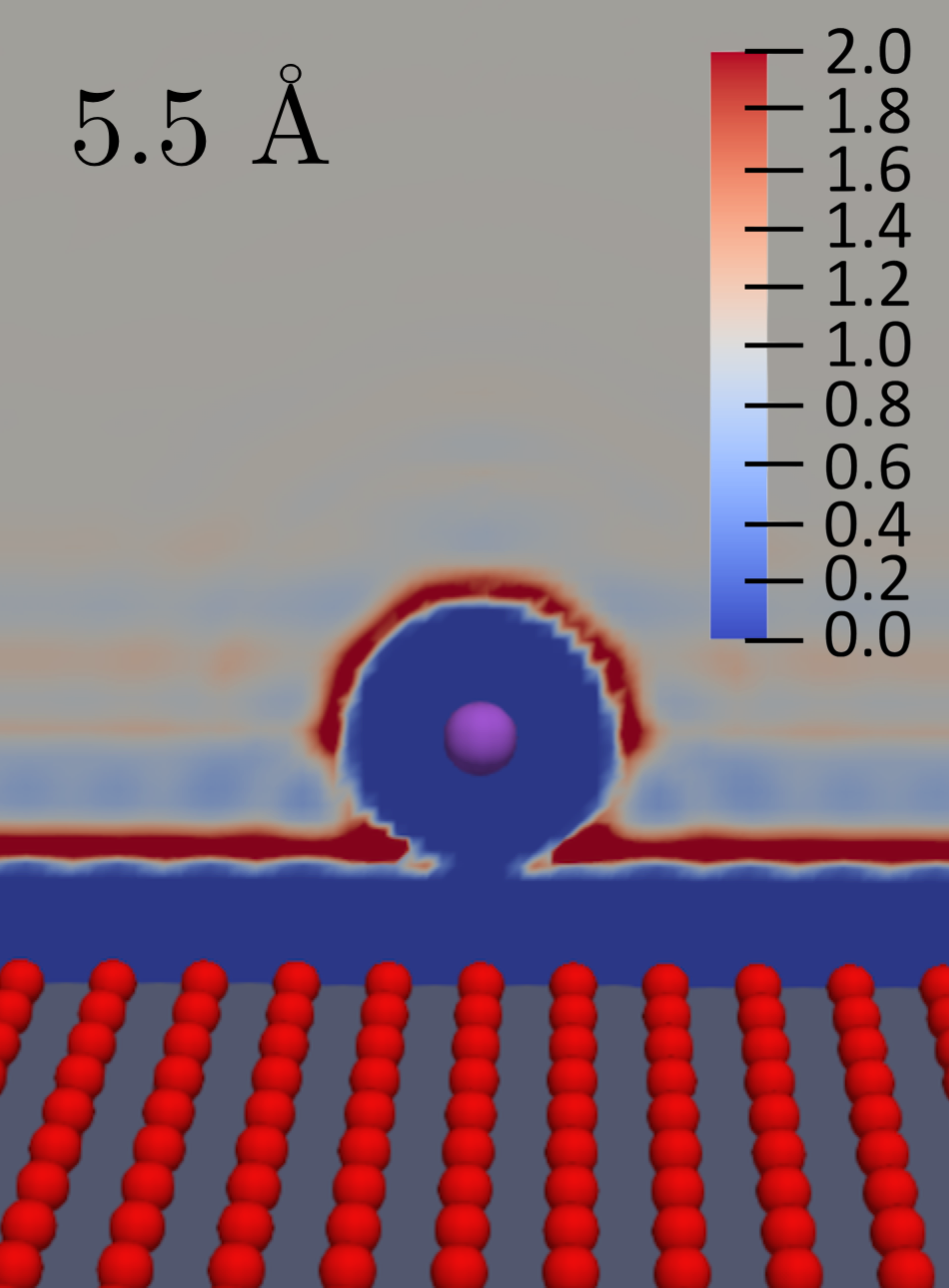}\tabularnewline
\includegraphics[width=0.2\textwidth]{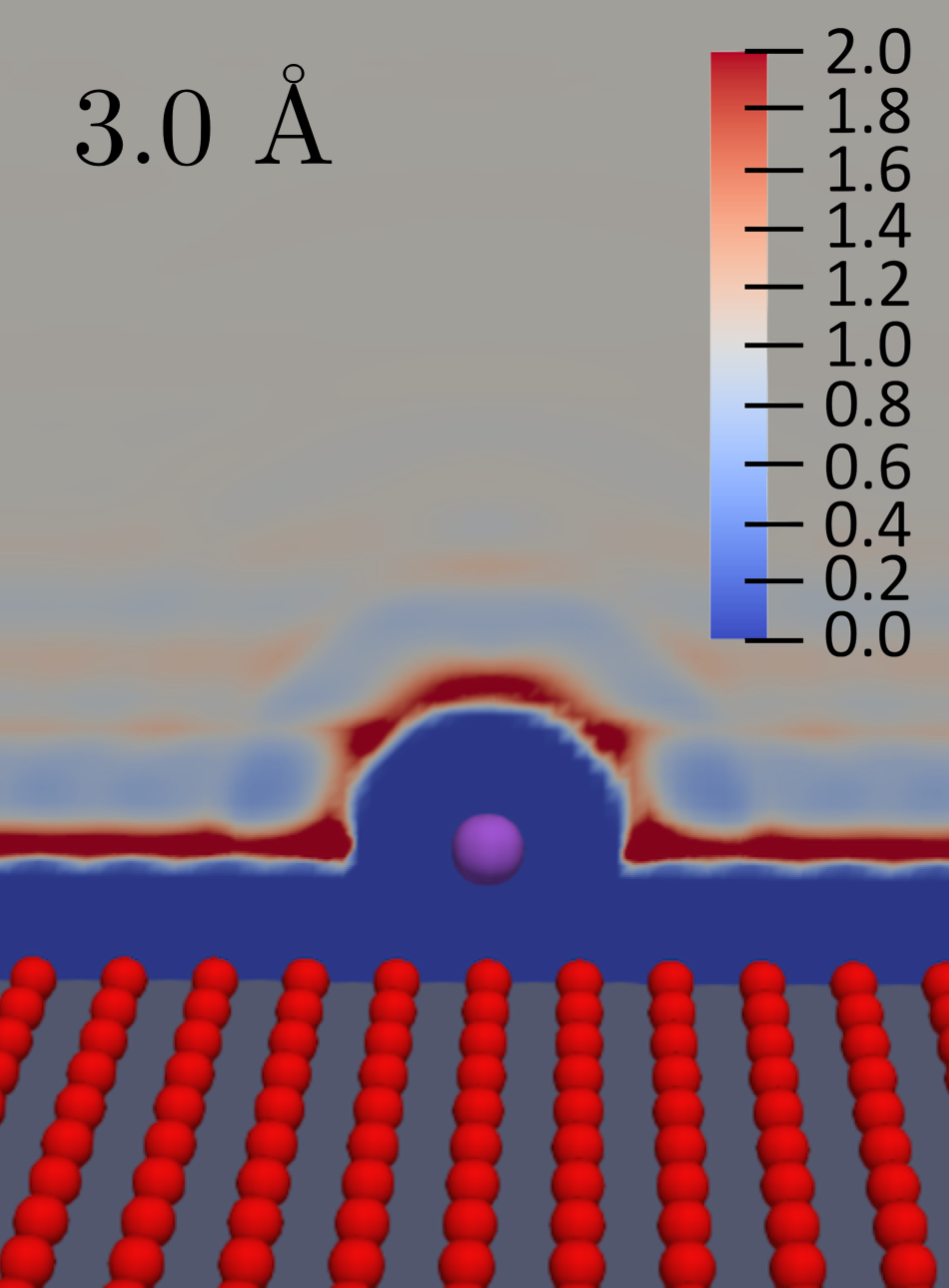} & \includegraphics[width=0.2\textwidth]{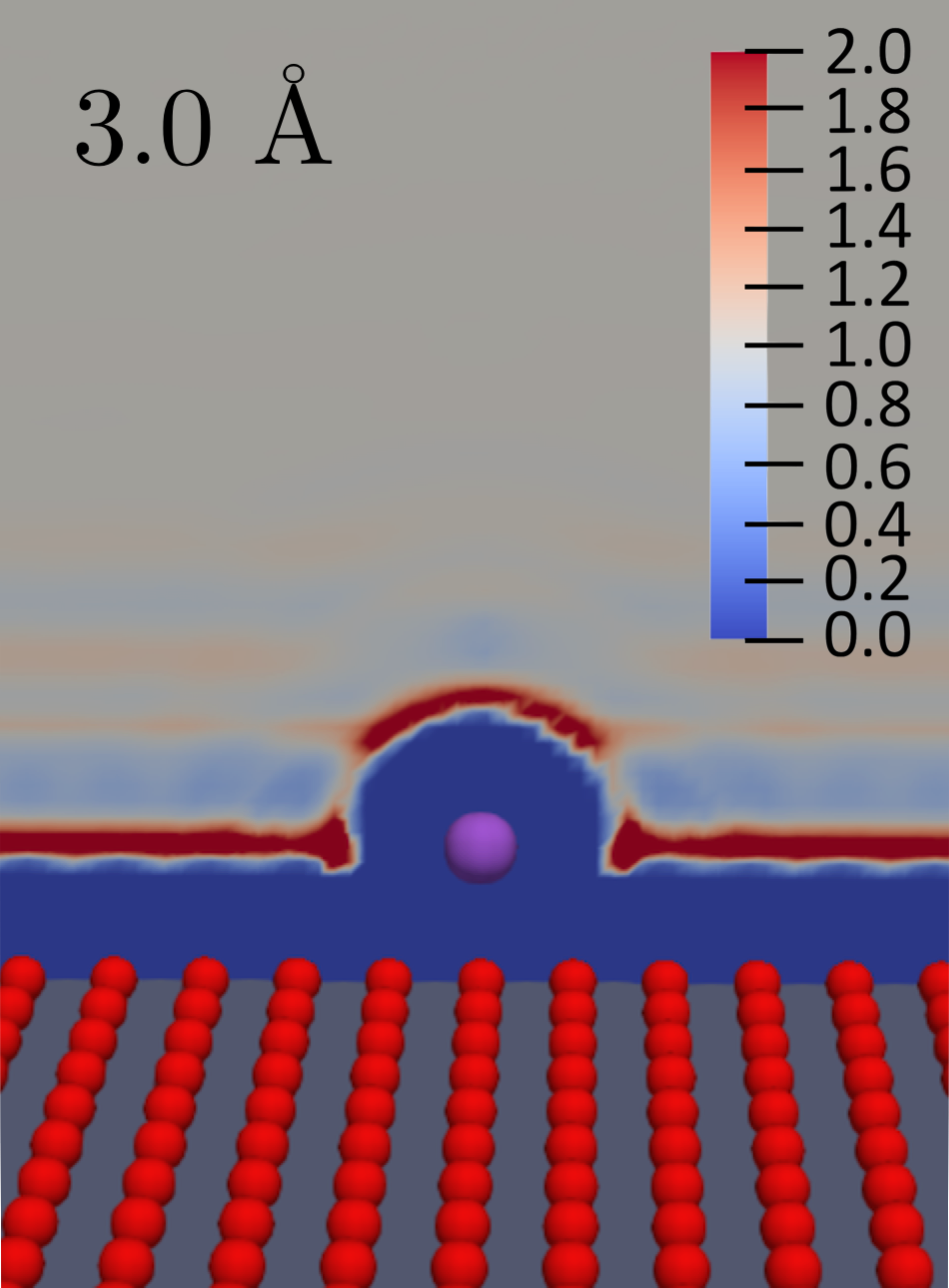}\tabularnewline
\end{tabular}\caption{Slices of the solvent density for various values of the distance $z$ from the wall. The neutral
solute data are shown is in the left-hand column, while the cation data are shown in the right-hand column. \label{FIG:denistywheiniongetscloser}}
\end{figure*}

We observe a decrease in the reorganization free energy as the solute
approaches the plane. We can rationalize this observation by realizing
that the wall truncates the solvation shell around the solute. This
effect is illustrated in Fig.\ \ref{FIG:denistywheiniongetscloser}
which shows slices  of the  density profile around the neutral (left column) and
charged (right column) solutes for different values of $z$. As the
solute approaches the wall, there are fewer solvent molecules to rearrange
when passing from one equilibrium solvation state to the other. This
reduces the cost of the reorganization and explains the decrease of
the free energy curves for small $z$. In the limit of total confinement
 the reorganization free energy would vanish.
The upper panel of Fig.\ \ref{FIG:lamda_wall} shows that reorganization
free energy of the charged solute exhibits a maximum around $5.5\ \textrm{\AA}$.
We rationalize this effect by decomposing $\lambda_{1}$ into its
components $F_{1}[\rho_{1}]$ and $F_{1}[\rho_{0}]\ $ in the lower 
panel of Fig.\ \ref{FIG:lamda_wall}. While $F_{1}[\rho_{0}]\ $ exhibits
a  marked maximum around $5\ \textrm{\AA}$, $F_{1}[\rho_{1}]$
has a maximum around $3.4\ \textrm{\AA}$ which is  flatter. Their
difference consequently gives rise to the oscillatory behavior of $\lambda_{1}$
around $5.5\ \textrm{\AA}$.

The first solvation shell of the neutral solute at $5.5\ \textrm{\AA}$
is in contact with the first fluid layer adsorbed on the wall. When
the solute gets closer, the solvation shell is reduced. This reduces
the unfavorable electrostatic term. It also decreases the cavity term
which measures the cost of expelling the solvent from the region around the
solute. This explains the 
maximum of $F_{1}[\rho_{0}].$ Considering the density for the charged
solute, the ``contact'' between the solvation shell of the solute
and the fluid layer adsorbed on the wall is also found around $5.5\ \textrm{\AA}$.
However, for the cation the truncation of the solvation shell decreases
both the favorable electrostatic term and the unfavorable cavity term.
This could explain why $F_{1}[\rho_{1}]$ is rather flat compared
to $F_{0}[\rho_{1}]$ and why the position of the maximum is shifted
to the left. For $z>10\ \textrm{\AA}$, the reorganization free energies
reach a plateau corresponding to the bulk value of Table \ref{tab:lamdaCl-2_DeltaF}.

The inset of panel a) shows the difference between $\lambda_1$ and $\lambda_0$ for the for the Cl$^0 \ \rightarrow$ Cl$^+$ ET, it presents a maximum at 5.5 $\textrm{\AA}$, \textit{i.e.} where the reorganization free energies also have a maximum. The difference never exceeds 6 kJ.mol$^{-1}$, so that Marcus' hypothesis is satisfied at all distances. This is clearly not the case for the for the Cl$^0 \ \rightarrow$ Cl$^-$ ET. Indeed, far from the wall $\lambda_0$ and $\lambda_1$ differ by 34 kJ.mol$^{-1}$ as reported in Table \ref{tab:lamdaCl-2_DeltaF}. However when the solute approaches the wall, this difference is reduced, indicating a decrease of the deviation the from linear response approximation. Again, this can be rationalized by the truncation of the solvation shell.

\begin{figure}[]
\centering{}%
\begin{tabular}{c}
\includegraphics[width=0.4\textwidth]{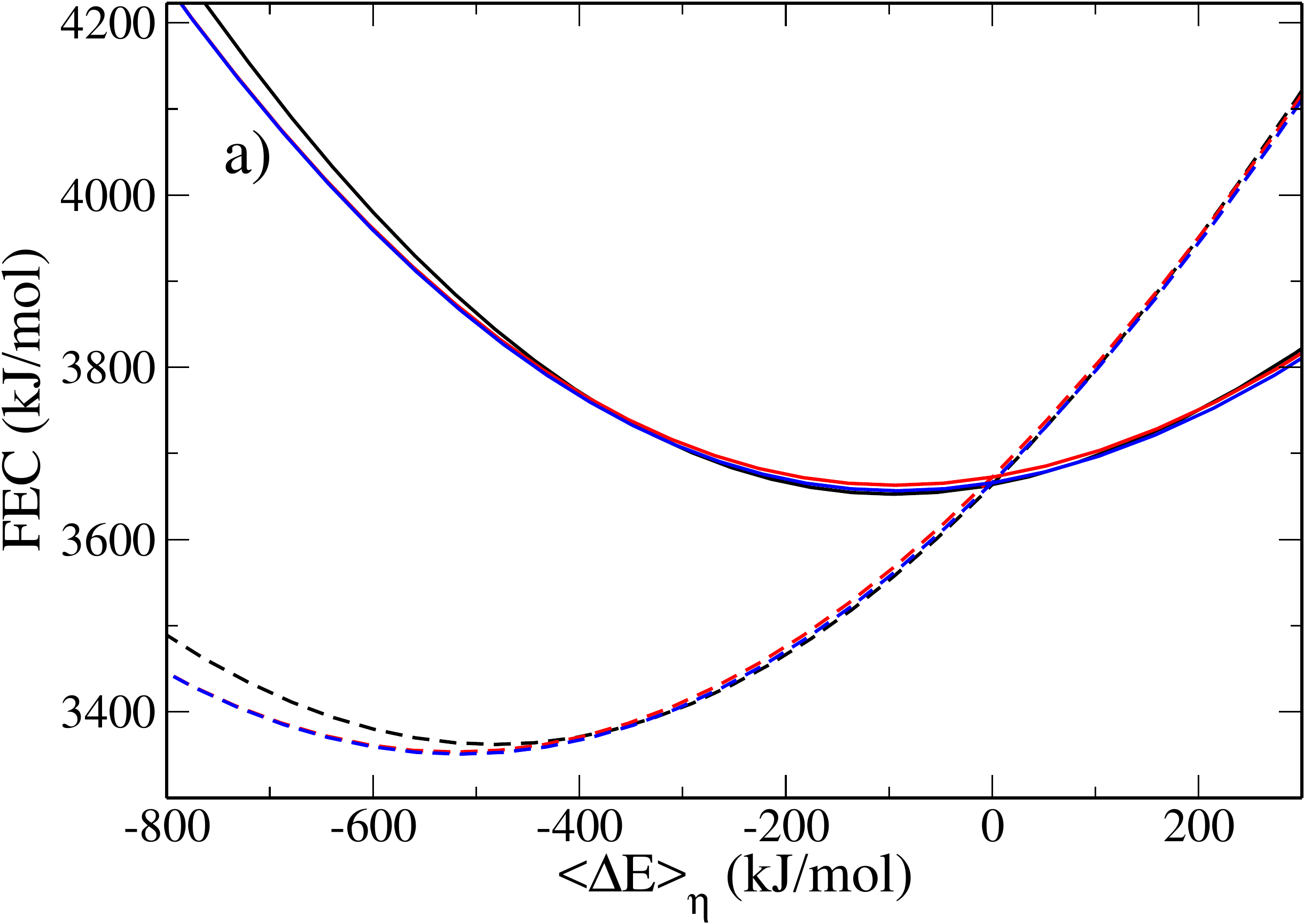} \tabularnewline
\includegraphics[width=0.4\textwidth]{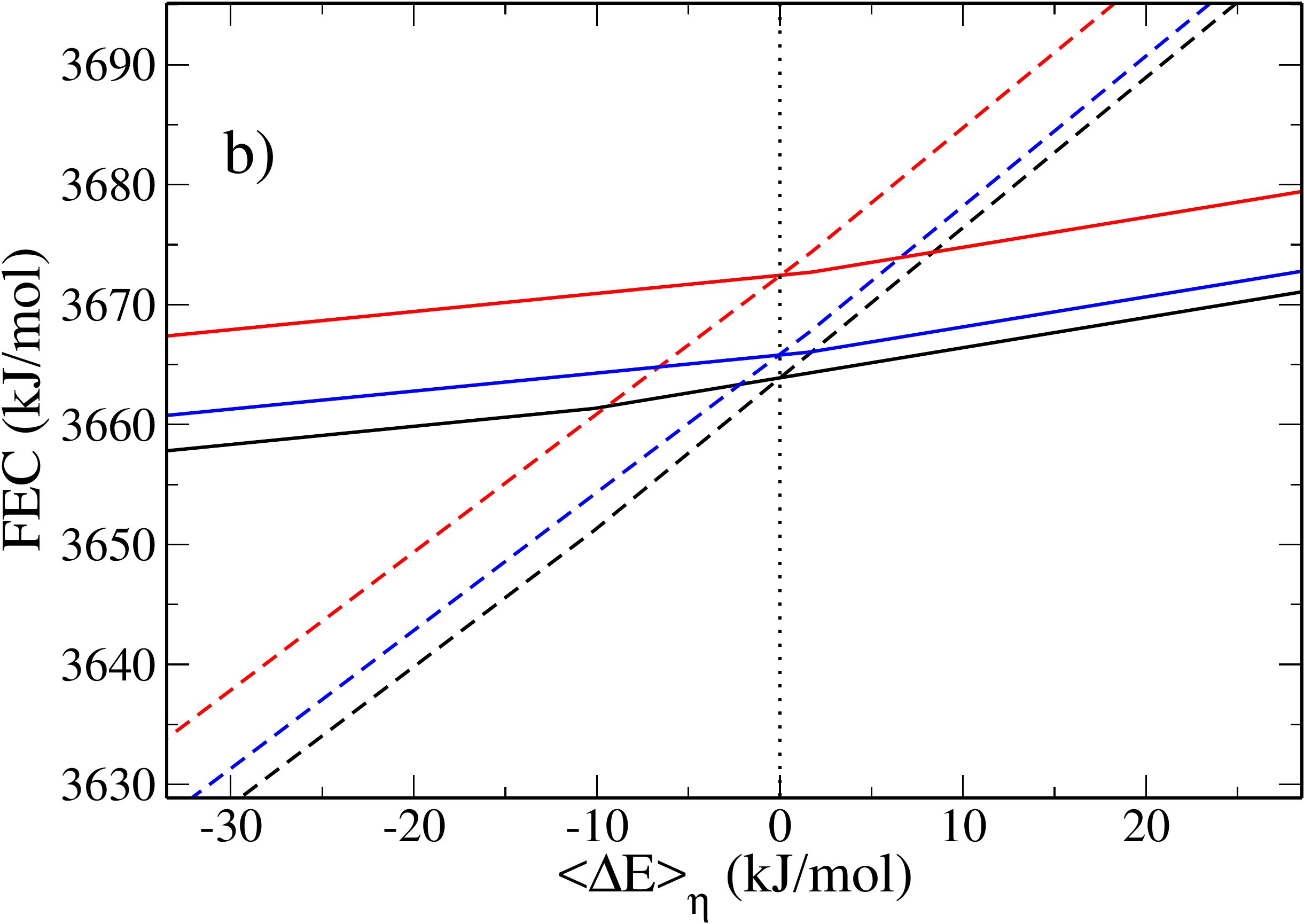}\tabularnewline
\includegraphics[width=0.4\textwidth]{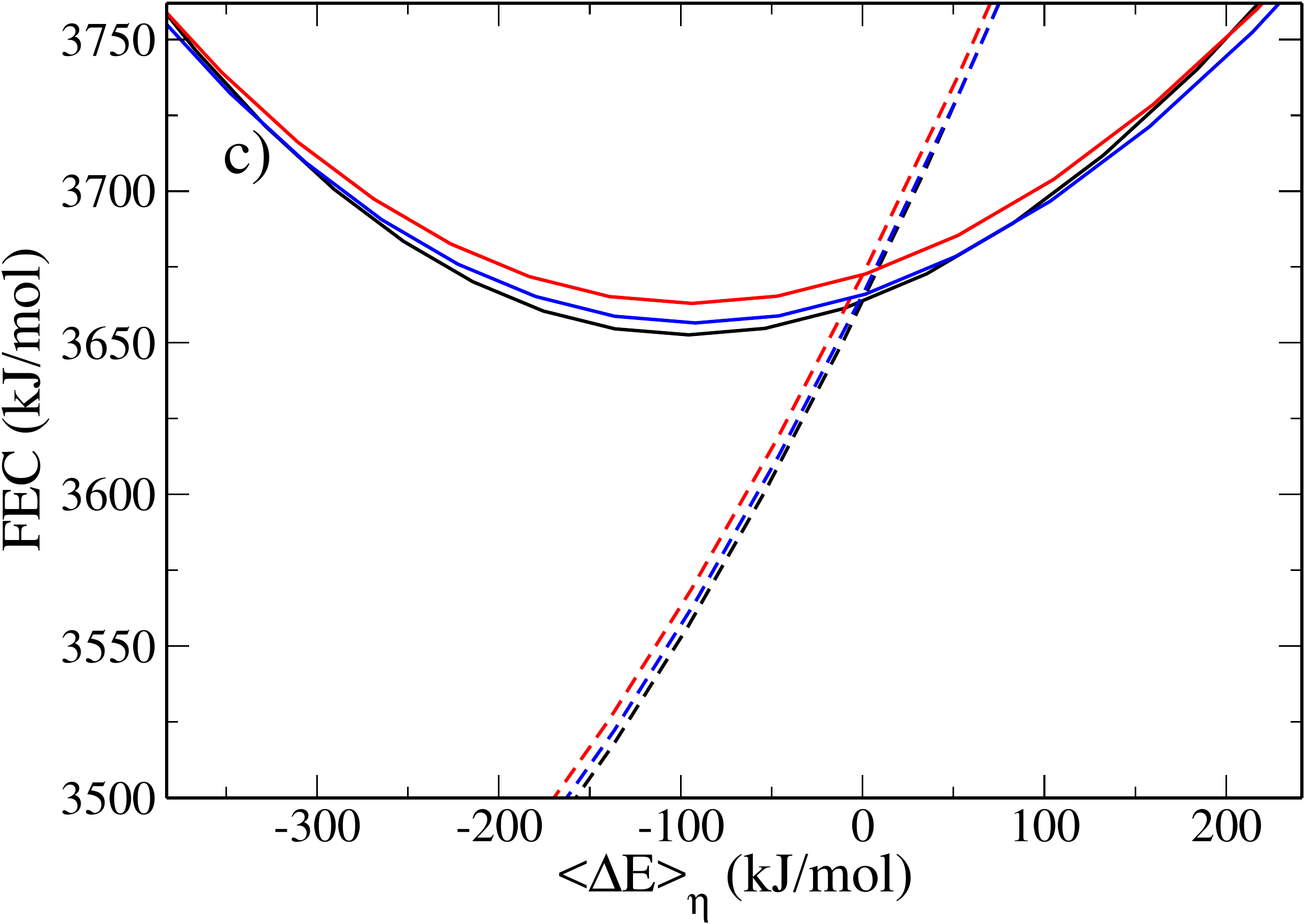} \tabularnewline
\includegraphics[width=0.4\textwidth]{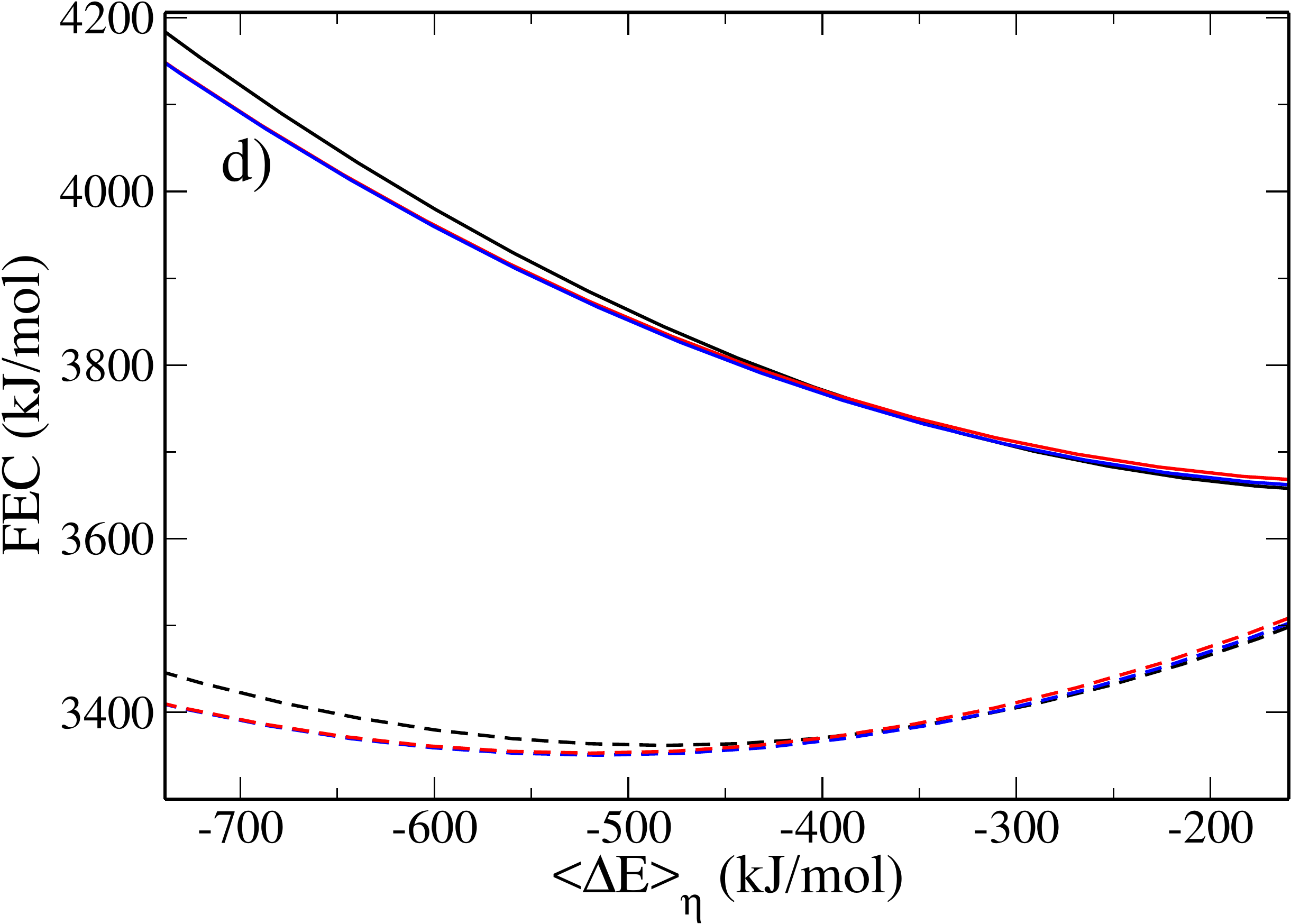}\tabularnewline
\end{tabular}\caption{a) Free energy  Cl$^{0}$ (full curve) and Cl$^{+}$ (dashed curve)
for various values of $z$. The black curves correspond to $z=3.0\ \textrm{\AA}$,
the red curves to $z=5.5\ \textrm{\AA}$ and the blue curves to $z=6.7\ \textrm{\AA}$.
\label{FIG:FECinconfimement}Panel b) is a zoom around $\left\langle \Delta E\right\rangle _{\eta}=0.0\ $kJ.mol$^{-1}$,
represented by a dotted lined. The c) and d) panels show 
zooms around the minimum of the atom and cation free energies respectively.}
\end{figure}

To illustrate the numerical efficiency of the method, we also computed
the FEC for various positions: $z=3.0\ \textrm{\AA}$
is in the region where the reorganization free energy decreases, $z=5.5\ \textrm{\AA}$
and $z=6.7\ \textrm{\AA}$ correspond to the first maximum and subsequent
local minimum in Fig.\ \ref{FIG:lamda_wall}. The FEC for the atom
and cation are presented in Fig.\ \ref{FIG:FECinconfimement}. Each
pair of curves cross at a point of vanishing vertical energy
gap, as expected. When the solute gets close to the wall the minimum
of the cation FEC is shifted towards positive values, which is
consistent with the above-mentioned truncation of the solvation shell.
Finally, the parabolas corresponding to $z=3.0\ \textrm{\AA}$ are wider
than those for higher values of $z$, which is consistent
with the smaller value of $\lambda$ close to the wall reported in
Fig.\ \ref{FIG:lamda_wall}. 

One of the advantages of MDFT is the possibility to split the
free energy into entropic, solute-solvent and solvent-solvent contributions
according to eq.\ref{eq:F=00003DFid+Fexc+Fext}:

\begin{align}
\label{eq:lambda_0splitted}
\lambda_{0}&=\left(F_{id}[\rho_{1}]-F_{id}[\rho_{0}]\right)+\left(F_{exc}[\rho_{1}]-F_{exc}[\rho_{0}]\right) \\ &+\int V_{0}(\bm{r},\bm{\Omega})(\rho_{1}-\rho_{0})(\bm{r},\bm{\Omega})d\bm{r}d\bm{\Omega} \nonumber
\end{align}
\begin{align}
 \label{eq:lamda1splitted}
\lambda_{1}&=\left(F_{id}[\rho_{0}]-F_{id}[\rho_{1}]\right)+\left(F_{exc}[\rho_{0}]-F_{exc}[\rho_{1}]\right)\\ &+\int V_{1}(\bm{r},\bm{\Omega})(\rho_{0}-\rho_{1})(\bm{r},\bm{\Omega})d\bm{r}d\bm{\Omega} \nonumber
\end{align}

\begin{figure}
\centering{}%
\begin{tabular}{c}
\includegraphics[width=0.4\textwidth]{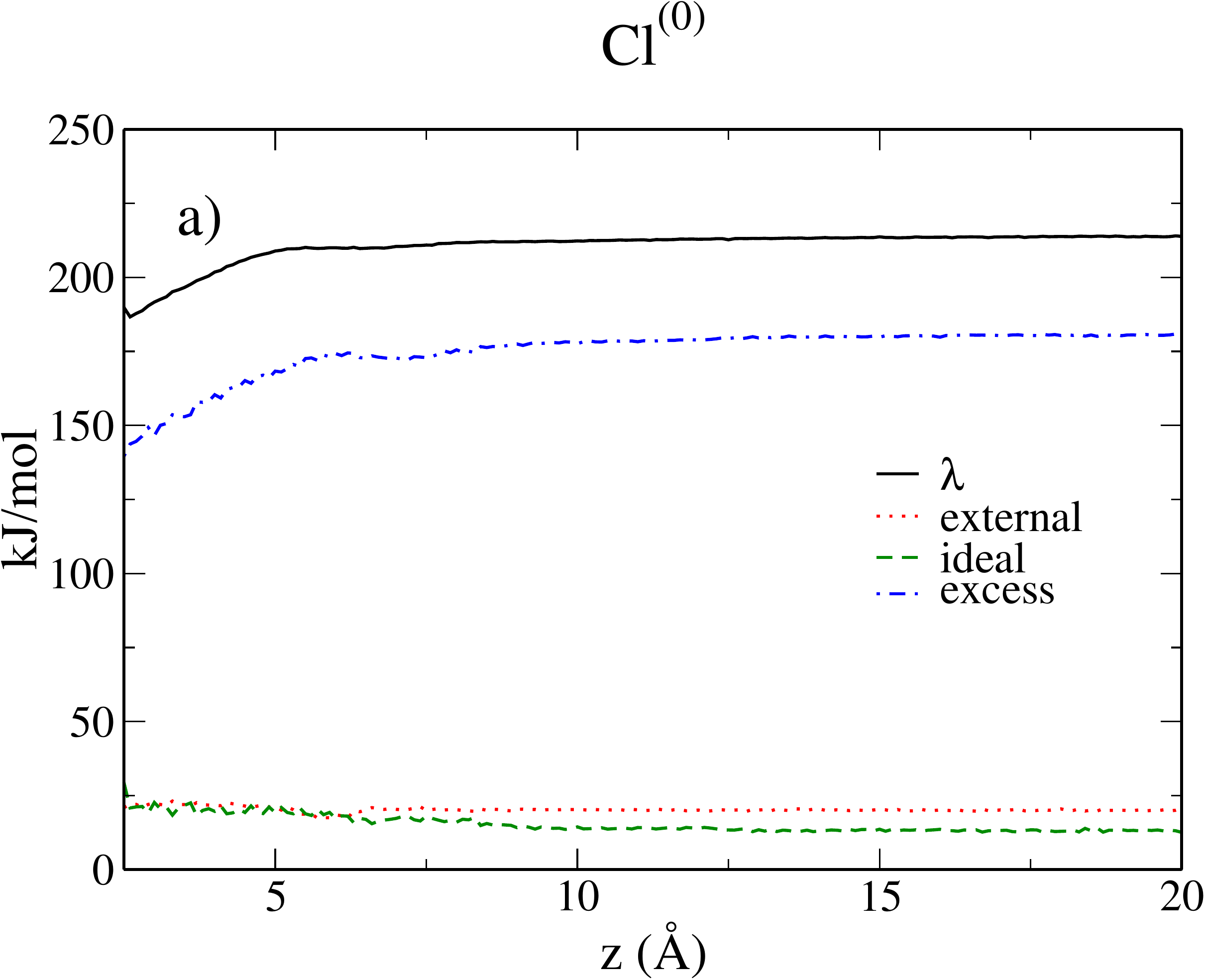} \tabularnewline
 \includegraphics[width=0.4\textwidth]{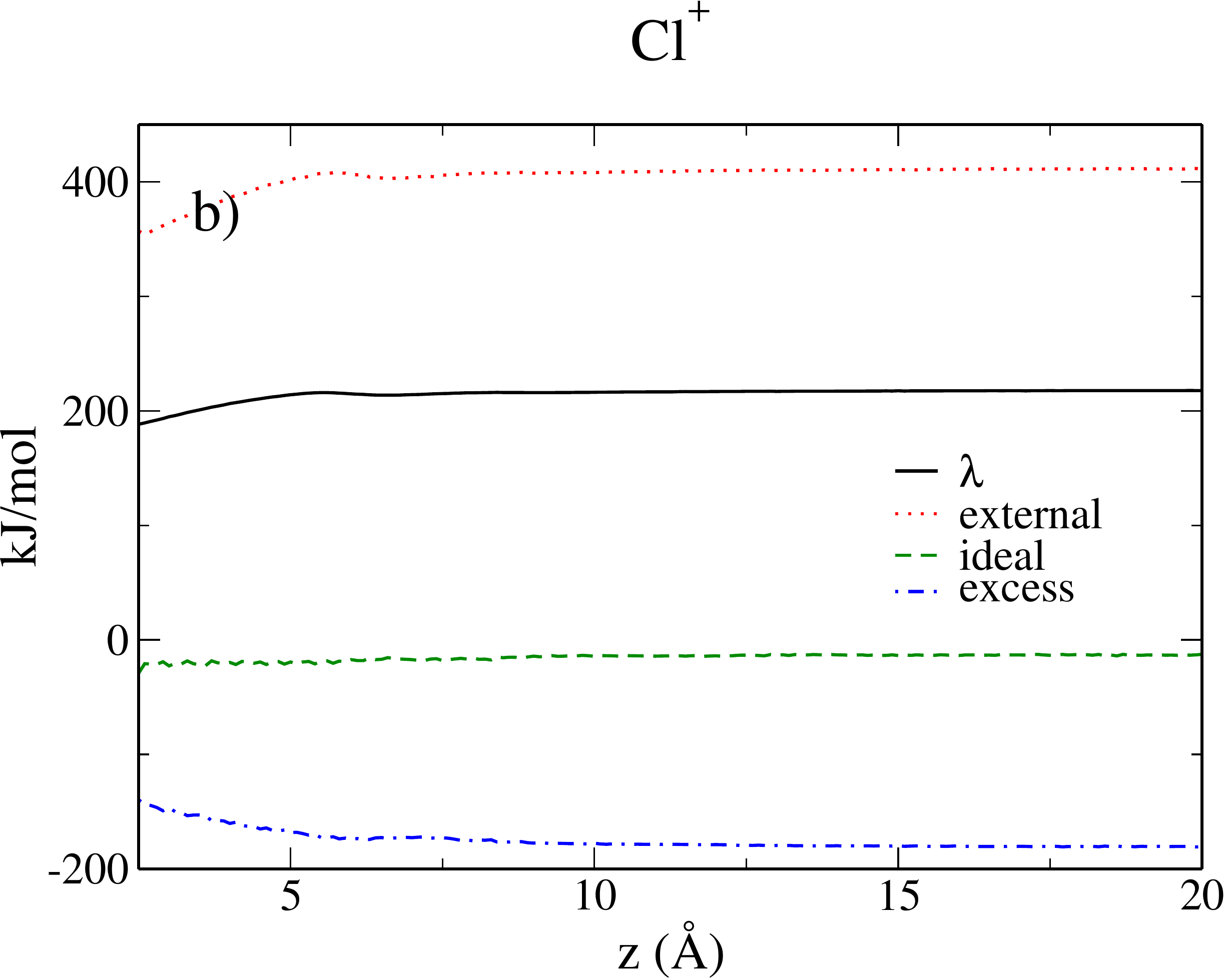}\tabularnewline
\end{tabular}\caption{Reorganization free energies and their different contributions for
Cl$^{(0)}$ (a) and Cl$^{+}$ (b) as a function of the distance from 
the wall computed as computed within MDFT. The reorganization free energy is in black,
the ideal term is in green, the excess term in blue and the external
term in red. \label{FIG:lambda_contrib}}
\end{figure}

Fig.\ref{FIG:lambda_contrib} shows the various contributions to
the reorganization free energy for the neutral and the charged solutes.
To our knowledge, this is the first time that such a decomposition of the reorganization free energy is reported.

A first conclusion emerging from eqs. \ref{eq:lambda_0splitted},
\ref{eq:lamda1splitted}, \ref{eq:Fid} and \ref{eq:Fexc def} is
that the ideal and  excess contributions are exactly opposite for
the neutral and the charged solutes. For both solutes, the ideal term
due to the entropic contribution remains quite small and hardly varies 
with the distance from the electrode.

For the neutral solute the external contribution is small due to the absence of electrostatic interactions and more
than 80\% of the reorganization free energy is due to the excess term,
\textit{i.e. }the solvent-solvent contribution. In contrast, for the
charged solute the main contribution is due to the electrostatic interaction
between the solute and the solvent, which is roughly twice 
 in absolute value than the solvent-solvent term.  Even if we 
already know from the previous subsection that the Cl$^0\rightarrow$ Cl$^+$
 transfer does satisfy the linear response approximation, \textit{i.e} $\lambda_0=\lambda_1$, 
 it is fascinating to observe the compensation of the three contributing terms resulting in this equality.
 When the solutes approaches the wall, the linear response approximation gets even better as evidenced in
  Fig.\ \ref{FIG:lamda_wall} where the curves of $\lambda_0$ and $\lambda_1$ converge.
 This study also illustrates the interest of MDFT not only to compute
the relevant free energies, but also to understand the various contributions
to the free energy.

\section{Conclusion\label{sec:Conclusions}}

Marcus theory plays a crucial role in the study of ET reactions.
This explains why its validity has been investigated extensively using molecular
dynamics simulation. However, MD remains computationally very demanding, and has so far been essentially limited to simple systems. Molecular density
functional theory has been proposed as an alternative to study
solvation because it is computationally  much faster, while retainning a molecular description
of the solvent. In the  present paper, we develop tools to use MDFT to
study electron transfer reactions in water using MDFT. We have first derived
how to compute the relevant reaction coordinate: the average vertical
energy gap. We have also shown how to compute the free energy curves
and the reorganization free energies.

We examined the validity of the approach by studying simple solutes,
namely the ET reactions between Cl$^{0}$, Cl$^{-}$ and Cl$^{+}$
modeled by a single Lennard Jones site and a point charge. We found
a good agreement between the results obtained by MDFT and corresponding 
MD simulations. We confirmed the effect reported by Hartnig \textit{et
al}., that the ET between neutral and positive solutes is
well described by Marcus theory, but not in the case
of the transfer between the neutral atom and the anion. 

We finally illustrated the potentiality of the method by tackling a more
challenging system. We investigated the effect of the presence of a solid/solvent interface 
on the reorganization free energy, using a model system composed of
an atomistically resolved neutral wall which is approached by the solute along the axis perpendicular to the wall. We
computed the reorganization free energy for both  neutral and charged
states and found that they exhibit similar features. The reorganization
free energy remains constant when the solute is far from the wall.
As it approaches the wall, it exhibits oscillations before decreasing.
We rationalized this behavior by considering the evolution of the
solvation shell: close to the wall, there is
less solvent to reorganize in the first solvation shell, thereby reducing
the free energy cost.

This work is a first step towards the study of ET reaction in water
and at electrode/water interfaces based on MDFT. The solvent effect sometimes called outer-sphere
contribution is not the only mechanism playing a role in the ET reaction.
The rearrangement of the electron cloud of the solute entering the
so-called inner-sphere contribution may also play an important role.
This effect is well taken into account in QM/MM calculation.
There are mainly two approaches to deal with the MM part in such calculations.
The first one is to use MD, which takes into account the molecular nature of the solvent, but remains computationally costly.
The second one is to use PCM-like models in which the solvent is
described as a dielectric continuum.  This approach neglects the molecular nature of solvent. 
As a consequence, it always assumes the validity of the linear response approximation and cannot properly describe systems violating Marcus theory.
The strength of this  method is its numerical efficiency: Calculations are almost instantaneous.
MDFT is thus a promising alternative to those two approaches to account for solvation in QM calculations: Even if it is computationally more demanding than PCM
 its computational cost  remains negligible compared to the cost of the QM calculation while its accuracy is comparable to MD.  
To that end, we are currently working on  coupling MDFT with electronic
structure calculations such as electronic density functional theory.
We also wish to develop
a framework allowing for the description of the polarizability of the
wall to describe electrodes at fixed electrode potential and study
electrochemical reactions. 
These two objective are  currently under investigation, in a attempt 
to develop a computationally efficient MDFT toolbox to tackle ET reactions.

\section*{Conflicts of interest}
There are no conflicts to declare.

\section*{Acknowledgments}

The authors acknowledge Luc Belloni for his precious inputs to MDFT.
The authors are also  grateful to  Jean-Pierre Hansen for his careful reading
of the manuscript.
B.R. acknowledges financial support from the French Agence Nationale
de la Recherche (ANR) under Grant No. ANR-15-CE09-0013 and from the
Ville de Paris (Emergences, project Blue Energy). This project has
received funding from the European Research Council (ERC) under the
European Union\textquoteright s Horizon 2020 research and innovation
programme (grant agreement No. 771294). This work was supported by
the Energy oriented Centre of Excellence (EoCoE), Grant Agreement
No. 676629, funded within the Horizon 2020 framework of the European
Union.

\appendix

\section{Proof that there is a one-to-one mapping between $\rho_{\eta}$ and $\left\langle \Delta E\right\rangle _{\eta}$ \label{App:equivalence_rhoeta_deltaEeta}}

A straightforward consequence of 
eq.\ref{eq:DeltaEetaapp} is that the average vertical energy gap is
uniquely defined by the density field. Following Mermin and Evans \citep{mermin_thermal_1965, evans_nature_1979},
we  proceed by \textit{reductio ad absurdum }to show that the
average vertical energy gap uniquely determines the external potential
and thus the density.\textit{ }Let us assume there exist two potentials
$V_{\eta}$ and $V_{\eta^{\prime}}$ with $\eta\neq\eta^{\prime}$
giving rise to the same gap \textit{i.e $\left\langle \Delta E\right\rangle _{\eta^{\prime}}=\left\langle \Delta E\right\rangle _{\eta}$.
}From the expression of the probability distribution in eq.\ref{eq:feta}
and as stated in Appendix 1 of Evans's article \citep{evans_nature_1979},
$V_{\eta}\neq V_{\eta^{\prime}}$ implies $f_{\eta}\neq f_{\eta^{\prime}}$.
From the variational principle of the grand potential we have
\begin{align}
\Theta_{\eta} & =  \text{Tr}\left[f_{\eta}\left(H_{\eta}-\mu N+k_{B}T\ln f_{\eta}\right)\right]\label{eq:Omegaeta=00003D...}\\
 & <  \text{Tr}\left[f_{\eta^{\prime}}\left(H_{\eta}-\mu N+k_{B}T\ln f_{\eta^{\prime}}\right)\right]\nonumber \\
 & <  \Theta_{\eta^{\prime}}+\left(\eta-\eta^{\prime}\right)\iint\rho_{\eta^{\prime}}(\bm{r},\bm{\Omega})\left[V_{1}(\bm{r},\bm{\Omega})-V_{0}(\bm{r},\bm{\Omega})\right]d\bm{r}d\bm{\Omega}.\nonumber 
\end{align}

By inverting the primed and unprimed quantities we get
\begin{equation}
\Theta_{\eta^{\prime}}<\Theta_{\eta}+\left(\eta^{\prime}-\eta\right)\iint\rho_{\eta}(\bm{r},\bm{\Omega})\left[V_{1}(\bm{r},\bm{\Omega})-V_{0}(\bm{r},\bm{\Omega})\right]d\bm{r}d\bm{\Omega}.\label{eq:Omegaetaprime}
\end{equation}

If we now sum eq.\ref{eq:Omegaeta=00003D...} and eq.\ref{eq:Omegaetaprime}
we arrive at
\begin{align}
\label{eq:absurde}\Theta_{\eta^{\prime}}+\Theta_{\eta}<&\Theta_{\eta^{\prime}}+\Theta_{\eta} \\ &+\left(\eta-\eta^{\prime}\right)\iint\left[\rho_{\eta^{\prime}}(\bm{r},\bm{\Omega})-\rho_{\eta}(\bm{r},\bm{\Omega})\right]\left[V_{1}(\bm{r},\bm{\Omega})-V_{0}(\bm{r},\bm{\Omega})\right]d\bm{r}d\bm{\Omega} \nonumber.
\end{align}

The integral on the r.h.s of eq.\ref{eq:absurde} vanishes as a consequence 
of eq.\ref{eq:DeltaEetaapp} and the assumption that \textit{$\left\langle \Delta E\right\rangle _{\eta^{\prime}}=\left\langle \Delta E\right\rangle _{\eta}$}
leading to a contradiction. Consequently, for this family of external
potential $V_{\eta}$ there is a unique $\left\langle \Delta E\right\rangle _{\eta}$
which corresponds to a given probability distribution $f_{\eta}$.
We hence have a one to one mapping between all the following quantities
\begin{equation}
\eta\leftrightarrow V_{\eta}\leftrightarrow f_{\eta}\leftrightarrow\rho_{\eta}\leftrightarrow\left\langle \Delta E\right\rangle _{\eta}
\end{equation}
where $\leftrightarrow$ denotes a one-to one-mapping. The bijections
between the three quantities $V$, $f$ and $\rho$ are always true
in the cDFT formalism \citep{evans_nature_1979} while the one involving
$\eta$ and $V_{\eta}$ is true within the class of potentials we
have chosen.

Because there is a bijection between $\left\langle \Delta E\right\rangle _{\eta}$
and a probability distribution, then the free energy of any state
uniquely depends on $\left\langle \Delta E\right\rangle _{\eta}$.
To express the free energy as a function of $\left\langle \Delta E\right\rangle _{\eta}$,
it is sufficient to take advantage of the one to one mapping between
$\rho_{\eta}$ and $\left\langle \Delta E\right\rangle _{\eta}$,
to obtain the expression of eq.\ref{eq:F0=00005Beta=00005D_1}

It is worth noticing that we can actually define $F_{0}\left(\left\langle \Delta E\right\rangle _{\eta}\right)$
as the Legendre transform of $\Theta_{\eta}$ with respect to $\eta$,
as $\left\langle \Delta E\right\rangle _{\eta}$ is the conjugate
variable of $\eta$: 
\begin{align}
\frac{d\Theta_{\eta}}{d\eta}&=\frac{d(-k_{B}T\ln\Xi_{\eta})}{d\eta}=\frac{\text{Tr}\left[\left(V_{1}-V_{0}\right)e^{-\beta\left(H_{0}+\eta(V_{1}-V_{0})-\mu N\right)}\right]}{\text{Tr}\left[e^{-\beta\left(H_{0}+\eta(V_{1}-V_{0})-\mu N\right)}\right]} \nonumber \\
&=\left\langle V_{1}-V_{0}\right\rangle _{\eta}=\left\langle \Delta E\right\rangle _{\eta}.
\end{align}

Moreover,
\begin{align}
\frac{d^{2}\Theta_{\eta}}{d\eta^{2}} & =  -\beta\frac{\text{Tr}\left[\left(V_{1}-V_{0}\right)^{2}e^{-\beta\left(H_{0}+\eta(V_{1}-V_{0})-\mu N\right)}\right]}{\text{Tr}\left[e^{-\beta\left(H_{0}+\eta(V_{1}-V_{0})-\mu N\right)}\right]} \nonumber \\ &+\beta\frac{\text{Tr}\left[\left(V_{1}-V_{0}\right)e^{-\beta\left(H_{0}+\eta(V_{1}-V_{0})-\mu N\right)}\right]^{2}}{\text{Tr}\left[e^{-\beta\left(H_{0}+\eta(V_{1}-V_{0})-\mu N\right)}\right]^{2}}\nonumber \\
 & =  -\beta\left(\left\langle \left(V_{1}-V_{0}\right)^{2}\right\rangle _{\eta}-\left(\left\langle V_{1}-V_{0}\right\rangle _{\eta}\right)^{2}\right).\label{eq:}
\end{align}

Therefore the second derivative of $\Theta_{\eta}$ is negative, so
that $\Theta_{\eta}$ is convex and its Legendre transform exists.
We can hence define the Legendre transform of $F_{\eta}$ by 
\begin{align}
F_{\eta}^{\star}\left(\left\langle \Delta E\right\rangle _{\eta}\right) & =  F_{\eta}[\rho_{\eta}]-\eta\left\langle \Delta E\right\rangle _{\eta}\nonumber\\
 & =  F_{\eta}[\rho_{\eta}]-\iint\rho_{\eta}(\bm{r},\bm{\Omega})(V_{\eta}(\bm{r},\bm{\Omega})-V_{0}(\bm{r},\bm{\Omega}))d\bm{r}d\bm{\Omega}\nonumber \\
 & =  F_{id}[\rho_{\eta}]+F_{exc}[\rho_{\eta}]+\iint\rho_{\eta}(\bm{r},\bm{\Omega})V_{0}(\bm{r},\bm{\Omega}))d\bm{r}d\bm{\Omega}\nonumber \\
 & =  F_{0}\left[\rho_{\eta}\right] \label{eq:Legendre} 
\end{align}
where we split $F_{\eta}$ into the sum of its three components as in
eq.\ref{eq:F=00003DFid+Fexc+Fext} and use the expression of $V_{\eta}$ in eq.\ref{eq:Veta-1-Appendix}. While the linear parametrization
of the external potential is the only one allowing to define $F_{0}\left(\left\langle \Delta E\right\rangle _{\eta}\right)$
as a Legendre transform, any parametrization leads the same expression
for $F_{0}\left(\left\langle \Delta E\right\rangle _{\eta}\right)$
and to the same FEC as demonstrated in Appendix \ref{App:Two-different-parametrization}.

\section{Two different parameterizations of $V_{\eta}$ lead to the same $F\left(\left\langle \Delta E\right\rangle _{\eta}\right)$\label{App:Two-different-parametrization}}

Let us consider the general parametrization for the interpolating potential,
\begin{equation}
V_{\eta}^{s}=V_{0}+s(\eta)\left(V_{1}-V_{0}\right)
\end{equation}
where $s$ is a strictly increasing continuous function with $s(0)=0$
and $s(1)=1$. We first show that any parametrization verifies the
properties demonstrated in Appendix \ref{App:equivalence_rhoeta_deltaEeta}. For any function $s$, let $\gamma,\delta\in[0,1]$
such as $\left\langle \Delta E\right\rangle _{\gamma}^{s}=\left\langle \Delta E\right\rangle _{\delta}^{s}$.
Using an argument identical to eq.\ref{eq:Omegaeta=00003D...} we
obtain
\begin{equation}
\Theta_{s(\gamma)}<\Theta_{s(\delta)}+\left[s(\delta)-s(\gamma)\right]\iint\rho_{\gamma}^{s}(\bm{r},\bm{\Omega})\left[V_{1}(\bm{r},\bm{\Omega})-V_{0}(\bm{r},\bm{\Omega})\right]d\bm{r}d\bm{\Omega}.
\end{equation}
Again the $\delta$ and $\gamma$ indexes can be interchanged to show
a one-to-one mapping between a value of the coupling parameter, the
external potential, the equilibrium probability distribution and the
equilibrium density. However this mapping now depends on the chosen
parametrization $s$,

\begin{equation}
\eta\overset{s}{\leftrightarrow}V_{\eta}^{s}\overset{s}{\leftrightarrow}f_{\eta}^{s}\overset{s}{\leftrightarrow}\rho_{\eta}^{s}\overset{s}{\leftrightarrow}\left\langle \Delta E\right\rangle _{\eta}^{s}.\label{eq:equivalence_s_eta}
\end{equation}

The mapping between $\left\langle \Delta E\right\rangle _{\eta}^{s}$
and $\rho_{\eta}^{s}$ leads to:
\begin{eqnarray}
F_{0}\left(\left\langle \Delta E\right\rangle _{\eta}^{s}\right) & = & F_{0}\left[\rho_{\eta}^{s}\right].\label{eq:F0_deltaetageneral}
\end{eqnarray}

This relation does not depend on the choice of the parametrization
$s$, but the values of $\left\langle \Delta E\right\rangle _{\eta}^{s}$
and $F_{0}\left[\rho_{\eta}^{s}\right]$ do. We now show that 
than any parametrization yields the same FEC.

Let us consider a strictly increasing continuous function with $s(0)=0$
and $s(1)=1$. The intermediate value theorem guarantees that $s$
takes all the value between 0 and 1,   once only. This is true for
all $s$ and in particular for the identity function corresponding
to the linear parametrization. This last property implies that for
all $s(\eta)\in[0,1]$, there exists a unique $\alpha\in[0,1]$ such
that
\begin{eqnarray*}
s(\eta) & = & \alpha\\
V_{\eta}^{s} & = & V_{\alpha}
\end{eqnarray*}

Since the potential uniquely defines the functional this implies the
same equality between all the properties in eq.\ref{eq:equivalence_s_eta},
i.e.
\begin{eqnarray*}
f_{\eta}^{s} & = & f_{\alpha}\\
\rho_{\eta}^{s} & = & \rho_{\alpha}\\
\left\langle \Delta E\right\rangle _{\eta}^{s} & = & \left\langle \Delta E\right\rangle _{\alpha}\\
F_{0}[\left\langle \Delta E\right\rangle _{\eta}^{s}] & = & F_{0}[\left\langle \Delta E\right\rangle _{\alpha}]
\end{eqnarray*}

This completes the proof  that all parametrization of the intermediate
potential leads to the same FEC.

\section{Thermodynamic cycle proposed by Chong and Hirata\label{sec:Appendix-B:-Thermodynamic}}

\begin{figure}
\begin{centering}
\includegraphics[width=0.4\textwidth]{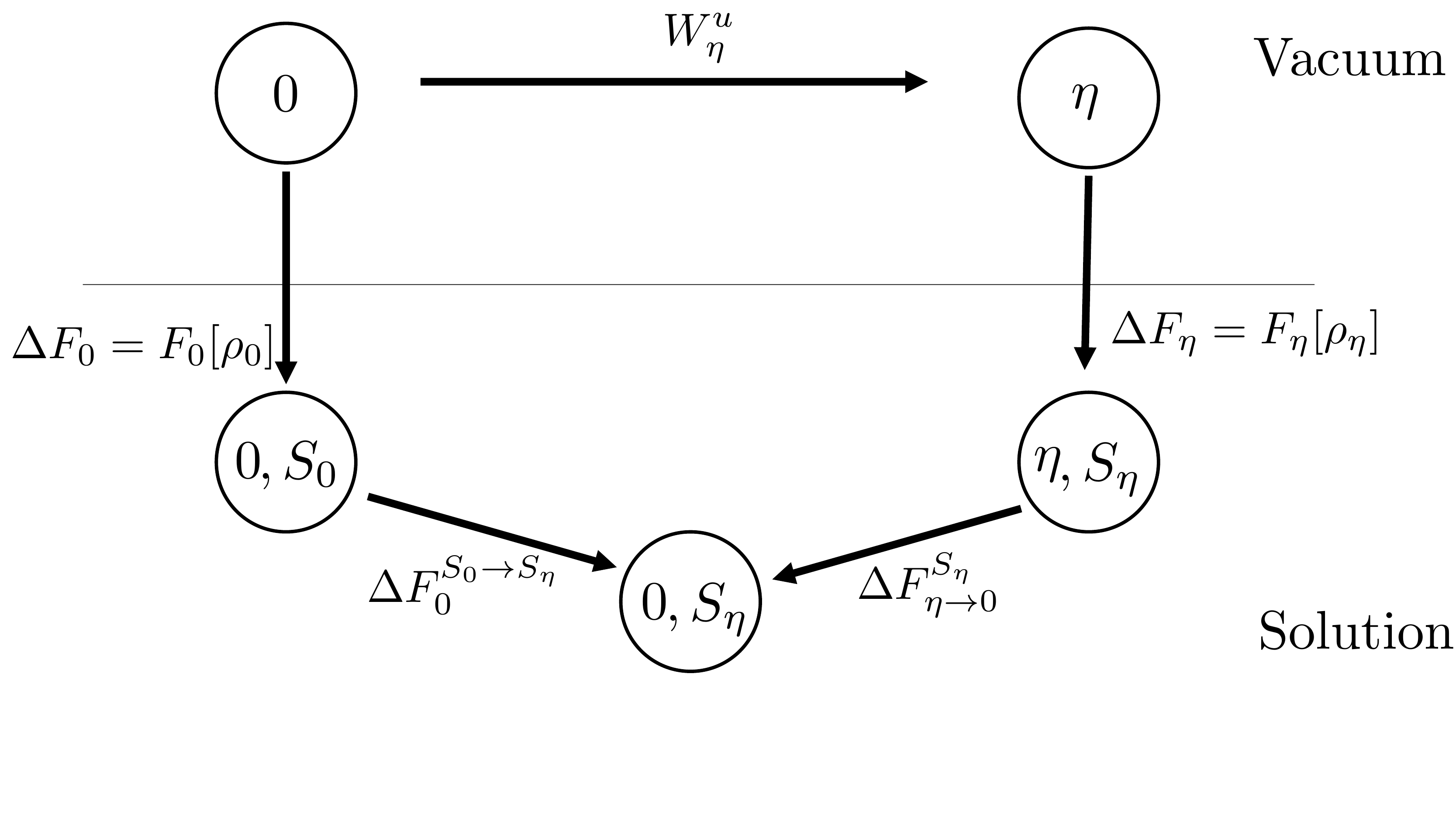}
\par\end{centering}
\caption{Schematic representation of the thermodynamic cycle used to recover 
eq.\ref{eq:Hirata_PES_0}. 0 corresponds to the  state under consideration, while
$\eta$ corresponds to a fictitious solute which interacts with the
solvent via an external potential $V_{\eta}$. The solvation states
in equilibrium with states 0 and $\eta$ are respectively denoted
by $S_{0}$ and $S_{\eta}$. To compute the FEC we need to compute
$\Delta F_{0}^{S_{0}\rightarrow S_{\eta}}$, the free energy cost
to modify the solvent configuration around state 0 from $S_{0}$ to
$S_{\eta}$. \label{fig:HirataSceme}}
\end{figure}

Chong and Hirata proposed the thermodynamic cycle displayed in Fig.\ \ref{fig:HirataSceme}
where 0 and $\eta$ are solutes corresponding to external potentials
$V_{0}$ and $V_{\eta}$ in eq.\ref{eq:Veta-1-Appendix} \citep{chong_free_1996}.
The objective is to find the free energy cost to modify the equilibrium
solvent configuration around 0 into a solvent configuration which would
be in equilibrium with $\eta$, a quantity denoted by $\Delta F_{0}^{S_{0}\rightarrow S_{\eta}}$.
Starting from 0 in vacuum, it is transformed into $\eta$ spending a work
$W_{\eta}^{u}$. Then, $\eta$ is solvated in its equilibrium solvent
configuration $S_{\eta}$. This step corresponds to the solvation
free energy $\Delta F_{\eta}$. The fictitious solute is transformed
into 0 while the solvent configuration is frozen within $S_{\eta}$. The
free energy cost of this step $\Delta F_{\eta\rightarrow0}^{S_{\eta}}$
can be split into the sum of two terms. The first one is the reversible
work to transform the solute in vacuum: It is the opposite of $W_{\eta}^{u}$.
The second term is the work $W_{\eta}^{v}$ to transform the solute
against the field created by the solvent configuration $S_{\eta}$
which can be expressed using our previous notation as:
\begin{equation}
W_{\eta}^{v}=\left\langle V_{0}-V_{\eta}\right\rangle _{\eta}.
\end{equation}

The final quantity required to close the cycle is the free energy cost
to solvate 0 into its equilibrium solvent configuration. This correspond
to the solvation free energy of state 0, $\Delta F_{0}$. By closing
the cycle, Chong and Hirata obtain the following formula:

\begin{equation}
\Delta F_{0}^{S_{0}\rightarrow S_{\eta}}=\Delta F_{\eta}-\Delta F_{0}+\left\langle V_{0}-V_{\eta}\right\rangle _{\eta}.\label{eq:Hirata_PES_0}
\end{equation}

If we replace the solvation free energy by the functional of the present
work, eq.\ref{eq:Hirata_PES_0} becomes

\begin{eqnarray}
\Delta F_{0}^{S_{0}\rightarrow S_{\eta}} & = & F_{\eta}[\rho_{\eta}]-F_{0}[\rho_{0}]+\left\langle V_{0}-V_{\eta}\right\rangle _{\eta}\nonumber \\
 & = & F_{\eta}[\rho_{\eta}]-F_{0}[\rho_{0}]+\iint\rho_{\eta}(\bm{r},\bm{\Omega})(V_{0}(\bm{r},\bm{\Omega})-V_{\eta}(\bm{r},\bm{\Omega}))d\bm{r}d\bm{\Omega}\nonumber \\
 & = & F_{0}[\rho_{\eta}]-F_{0}[\rho_{0}]\label{eq:-1}
\end{eqnarray}
which is equivalent to our previous finding.

\section{Modification of Hartnig and Koper's data to plot figure \ref{fig:Wcl}
\label{sec:Appendix-C:-Modification}}

In their paper, Hartnig and Koper represented their free energy
curves as  functions of a generalized order parameter defined as the
electrostatic interaction energy between a negative point charge at
the site of the solute and the solvent molecules \citep{hartnig_molecular_2001}.
Because they only considered solutes with a single Lennard-Jones site
which is kept unchanged during the ET the vertical energy gap is equal
to their order parameter for an anion and to its opposite for a cation.

They do not mention the use of any finite size effect corrections
while we we use that proposed by H\"unenberger \textit{et al} \citep{kastenholz_computation_2006,kastenholz_computation_2006-1}  which applies
to our case but also to MD simulations with Ewald electrostatics \citep{kastenholz_computation_2006}.
To convert their order parameter in vertical energy gap 
we thus i) multiply it by -1 in the case of the cation, ii) apply
the above mentioned electrostatic corrections with the box length
parameter of $L=24.83\ \textrm{\AA}$ reported in their paper.

Finally, Hartnig and Koper shifted all the FEC such that the minimum
of each curve is equal to 0. As a consequence their curves do not
cross for $\left\langle \Delta E\right\rangle _{\eta}=0$, as should
be the case by definition. Because they did not report the values
of the solvation free energy, or equivalently the values of the shift
applied to each curve, we decided to freely shift vertically the curves
corresponding to the atom in order to have the minimum of the curves
agree with the value predicted by MDFT. The MD curve for the ion has
been subsequently shifted vertically to fulfill the zero gap condition.

\bibliography{ET} %You need to replace "rsc" on this line with the name of your .bib file
\bibliographystyle{apsrev} %the RSC's .bst file

\end{document}